\documentclass[11pt,a4paper]{article}
\pdfoutput=1
\usepackage{jheppub}
\usepackage{rotating}
\usepackage{array}
\usepackage{amsmath}
\usepackage[normalem]{ulem}
\usepackage{slashed}
\usepackage{booktabs}
\usepackage[pdftex,table]{xcolor}
\usepackage{units}
\usepackage{enumerate}
\usepackage{upgreek}
\usepackage{xcolor}
\usepackage{xspace}
\usepackage{textcomp}

\newcommand{\mrhod}{m_{\rho_\mathrm{d}}}
\newcommand{\mpid}{m_{\pi_\mathrm{d}}}

\newcommand{\rhod}{{\rho_\mathrm{d}}}
\newcommand{\rhodzero}{{\rho^0_\mathrm{d}}}
\def \belletwo {Belle\,II\xspace}

\preprint{
\begin{flushright}
P3H-22-033 \\ TTK-22-12\\ FERMILAB-PUB-22-140-T\\ TTP22-019\\ DESY-22-047
\end{flushright}}
\arxivnumber{}

\title{Forecasting dark showers at \belletwo}

\author[1,2]{Elias Bernreuther,}
\author[1]{Kai B\"ose,}
\author[3]{Torben Ferber,}
\author[4,5]{Christopher Hearty,}
\author[1,6]{Felix~Kahlhoefer,}
\author[1]{Alessandro Morandini}
\author[7]{and Kai Schmidt-Hoberg}

\affiliation[1]{Institute for Theoretical Particle Physics and Cosmology (TTK), RWTH Aachen University, D-52056 Aachen, Germany}
\affiliation[2]{Fermi National Accelerator Laboratory, Batavia, IL 60510, USA}
\affiliation[3]{Institute for Experimental Particle Physics (ETP),  Karlsruhe Institute of Technology (KIT), D-76131 Karlsruhe, Germany}
\affiliation[4]{Department of Physics and Astronomy, University of British Columbia (UBC), Vancouver, British Columbia, V6T 1Z1 Canada}
\affiliation[5]{Institute of Particle Physics (Canada), Victoria, British Columbia V8W 2Y2, Canada}
\affiliation[6]{Institut f\"{u}r Theoretische Teilchenphysik, Karlsruhe Institute of Technology (KIT), 76128 Karlsruhe, Germany}
\affiliation[7]{Deutsches Elektronen-Synchrotron DESY, Notkestr.~85, 22607 Hamburg, Germany}

\emailAdd{ebernreu@fnal.gov}
\emailAdd{kai.boese@rwth-aachen.de}
\emailAdd{torben.ferber@kit.edu}
\emailAdd{hearty@physics.ubc.ca}
\emailAdd{kahlhoefer@physik.rwth-aachen.de}
\emailAdd{morandini@physik.rwth-aachen.de}
\emailAdd{kai.schmidt-hoberg@desy.de}

\abstract{
Dark showers from strongly interacting dark sectors that confine at the GeV scale can give rise to novel signatures at $e^+e^-$ colliders. In this work, we study the sensitivity of $B$ factory experiments to dark showers produced through an effective interaction arising from a heavy off-shell mediator. We show that a prospective search for displaced vertices from GeV-scale long-lived particles at \belletwo can improve the sensitivity to dark showers substantially compared to an existing search at BaBar. We compare the sensitivity of searches for displaced signals to searches for promptly produced resonances at BaBar and KLOE and calculate sensitivity projections for a single-photon search at \belletwo to invisible dark showers produced through an effective interaction. The underlying structure of the effective interaction can be resolved at higher-energy experiments, where the mediator can be produced on-shell. To study the resulting constraints, we update electroweak precision bounds on kinetically mixed $Z'$ bosons and reinterpret a search for low-mass di-muon resonances at LHCb in terms of dark showers. We find that LHCb and \belletwo are most sensitive to different particle decay lengths, underscoring the complementarity of LHC and intensity frontier experiments. 
}

\begin{document}

\maketitle

\section{Introduction}

The possibility that dark matter (DM) particles experience strong self-interactions was originally motivated by cosmology~\cite{Carlson:1992fn} and astrophysical observations~\cite{Spergel:1999mh}, but the implications for model building and laboratory experiments has continued to intrigue the particle physics community. Of particular interest is the possibility that DM particles form as bound states of a strongly interacting dark sector that experiences confinement at low energies~\cite{Strassler:2006im,Hochberg:2014kqa,Kribs:2016cew}. A promising strategy to explore this scenario is to try and produce dark sector states with sufficiently high energy that the fundamental constituents (i.e.\ dark quarks and gluons) are revealed and the details of hadronisation can be studied~\cite{Cohen:2017pzm,Pierce:2017taw,Beauchesne:2017yhh,Renner:2018fhh,Beauchesne:2019ato,Knapen:2021eip}.

The resulting dark shower can give rise to a wide range of exciting experimental observables. In the context of DM models, one expects a large part of the shower to remain invisible, as only a fraction of the dark mesons are able to decay into Standard Model (SM) particles. If these decays happen promptly and involve SM hadrons, the resulting signature is called a semi-visible jet~\cite{Cohen:2015toa}, which may be distinguished from conventional jets using a range of kinematic variables~\cite{Kar:2020bws,Cohen:2020afv,Beauchesne:2021qrw} and machine-learning techniques~\cite{Bernreuther:2020vhm} (see also the very recent CMS analysis in Ref.~\cite{CMS:2021dzg}).\footnote{For dark shower signatures that are associated with isotropic events instead of jets see Refs.~\cite{Strassler:2008bv,Knapen:2016hky,Costantino:2020msc,Cesarotti:2020uod,Barron:2021btf}.} If the decaying dark mesons are long-lived, experimental signals can be much more striking and may involve individual displaced vertices (DVs)~\cite{Bernreuther:2020xus} or entire emerging jets~\cite{Schwaller:2015gea,CMS:2018bvr}.

Although the LHC has established a strong programme to search for long-lived particles (LLPs)~\cite{Lee:2018pag,Alimena:2019zri}, the primary focus has been on LLPs with a mass well above the $B$ meson mass, and the sensitivity of current searches rapidly degrades for smaller masses~\cite{Bernreuther:2020xus}. Strongly interacting dark sectors, however, can be viable across a wide range of scales in the sense that the observed DM relic abundance can be reproduced and all astrophysical and laboratory constraints on the DM particles are satisfied~\cite{Berlin:2018tvf,Bernreuther:2019pfb}. This observation makes it an interesting and timely task to explore the sensitivity of particle physics experiments to strongly interacting dark sectors at the GeV scale and below.

In the present work we explore the sensitivity to dark showers of recently proposed searches for DVs at \belletwo~\cite{Duerr:2019dmv,Duerr:2020muu}, which benefit from the hermetic detector and optimised triggers. For this purpose we derive an effective description of dark shower production under the assumption that the particle mediating the interactions between the dark sector and the SM is heavy compared to \belletwo energies (see Refs.~\cite{Darme:2020ral,Boehm:2020wbt,Bertuzzo:2020rzo,Zhang:2021orr} for similar recent studies in different contexts). It turns out that in this case the dark shower production cross section can be expressed in terms of the mass and lifetime of the unstable dark mesons, which greatly simplifies the analysis and the presentation of our results. However, in contrast to the case where the mediating particle can be produced on-shell at \belletwo energies~\cite{Berlin:2018tvf}, the dark shower events do not contain mono-energetic photons that can be used for background suppression, which significantly degrades the sensitivity of \belletwo for the case that the dark shower remains fully invisible~\cite{Essig:2013vha}. Nevertheless, we find that existing and proposed triggers are sufficient to search for dark showers that produce DVs with great sensitivity.

To provide context for our sensitivity estimates, we consider a number of additional constraints. The same effective description can be used to reinterpret existing constraints from BaBar on exotic particles decaying promptly~\cite{BaBar:2014zli} or from a DV~\cite{BaBar:2015jvu} into a pair of charged leptons, as well as constraints on hadronic and muon resonances from KLOE~\cite{KLOE-2:2016ydq,KLOE-2:2018kqf}. Constraints from model-independent LLP searches at LHCb~\cite{LHCb:2020ysn} as well as from electroweak precision tests (EWPT), on the other hand, are sensitive to the detailed coupling structure of the underlying model. We explore these model-dependent constraints for several different settings and find an intriguing complementarity between the constraints from high and low energies.

The remainder of this work is structured as follows. In section~\ref{sec:eff} we introduce the model that we consider and derive the corresponding low-energy effective description of dark shower production. Section~\ref{sec:DV} then focuses on displaced vertices arising from such dark showers and determines the corresponding constraints from an existing BaBar search as well as the sensitivity of a proposed \belletwo search. In section~\ref{sec:lowe} we compare our results with additional constraints from low-energy experiments, which can be expressed in terms of the same effective parameters. Finally, section~\ref{sec:highe} considers complementary constraints from high-energy accelerators, which exhibit a more complicated dependence on the fundamental model parameters.

\section{Effective production of dark showers}
\label{sec:eff}

In this section we briefly describe the dark sector model underlying the dark shower signal discussed throughout this work and discuss the effective description of dark shower production at $B$ factory experiments that is applicable when the dark sector and SM particles are connected by a heavy mediator.

\subsection{Strongly interacting dark sector}

We consider the strongly interacting dark sector introduced in Ref.~\cite{Bernreuther:2019pfb} containing two flavours of dark quarks $q_\mathrm{d}$, which are SM singlets and transform in a hidden $SU(3)_\mathrm{d}$ gauge group. In addition, the two dark quark flavours have opposite charges $\pm1$ under a broken $U(1)'$ gauge group, which gives rise to a massive $Z'$ vector mediator.\footnote{Since we only consider vector-like couplings of the $Z'$, its mass can in principle be generated by the Stueckelberg mechanism. If it is instead generated by a dark Higgs mechanism, there may be additional collider signatures involving the dark Higgs boson that can be used to probe this model~\cite{Darme:2017glc,Duerr:2017uap}. The dark shower signature considered here will however not be affected.}
Hence, the dark quark Lagrangian takes the form
\begin{align}
\label{eq:lagrangian_darkquarks}
\mathcal{L}_\text{d} =  \sum_{i=1}^2 \overline{q}_{\mathrm{d},i} (i \slashed{D} - m_{q_\mathrm{d}}) q_{\mathrm{d},i} - e_\mathrm{d} Z^\prime_\mu \left(\overline{q}_{\mathrm{d},1}\gamma^\mu q_{\mathrm{d},1}- \overline{q}_{\mathrm{d},2}\gamma^\mu q_{\mathrm{d},2}\right) \; ,
\end{align}
with the dark quark mass $m_{q_\mathrm{d}}$ and the $U(1)'$ gauge coupling $e_\mathrm{d}$.
Below the scale $\Lambda_\mathrm{d}$, the dark sector confines, which leads to the formation of dark mesons. In particular, this confinement gives rise to three dark pions $\pi_\mathrm{d}^0$, $\pi_\mathrm{d}^+$, $\pi_\mathrm{d}^-$, with the $U(1)'$ charges $0$, $+2$ and $-2$, which are stable DM candidates and can arrive at the correct DM relic abundance via Boltzmann-suppressed annihilations into slightly heavier dark mesons with a rate that is exponentially sensitive to the mass splitting between the involved dark meson species.~\cite{DAgnolo:2015ujb, Li:2019ulz}. These other dark meson species are generically unstable~\cite{Berlin:2018tvf}. In particular, the neutral dark vector mesons $\rhodzero$ can decay by mixing with the $Z'$ mediator and thus give rise to visible signals in accelerator experiments.

The particle composition of the signal, as well as the production of dark showers, depend on the coupling structure of the $Z'$ to SM particles. Since we will study dark showers both at $e^+e^-$ and hadron colliders, it is necessary to specify the coupling of the $Z'$ to both SM quarks and leptons. Furthermore, we will be interested in the case where the $Z'$ couples much more strongly to dark quarks than to SM fermions, such that conventional constraints on $Z'$ bosons from di-lepton and di-jet resonance searches are suppressed. It is therefore well-motivated to consider a $Z'$ mediator with photon-like couplings induced by kinetic mixing with the SM hypercharge field~\cite{Babu:1997st}:
\begin{equation}
    \mathcal{L} \supset - \frac{\kappa}{2 c_\mathrm{w}} \hat{Z}^{\prime \mu \nu} \hat{B}_{\mu\nu} \; ,
\end{equation}
where $X^{\mu\nu} = \partial^\mu X^\nu - \partial^\nu X^\mu$ with  $X = \hat{Z}', \hat{B}$ is the field-strength tensor before diagonalisation, $c_\mathrm{w}$ denotes the cosine of the weak mixing angle and $\kappa$ the kinetic mixing parameter. 
After transforming to mass eigenstates with canonical kinetic terms, the interaction of the $Z'$ with SM fermions is given by
\begin{align}
\mathcal{L}_\text{int} = -\kappa e Z'_\mu \sum_{f} q_f \overline{f} \gamma^\mu f \; ,
\end{align}
where $q_f$ denotes the electric charge of the fermion $f$.

\subsection{Effective portal interaction}

Let us now turn to the production of dark sector states in collisions of SM fermions. If the centre-of-mass energy $\sqrt{s}$ is small compared to the mass of the $Z'$ mediator but large compared to the dark confinement scale $\Lambda_\mathrm{d}$, we can describe the interaction of the dark sector with SM particles in terms of the effective four-fermion operator
\begin{align}
\label{eq:smfermions_darkquarks_effective}
\mathcal{L}_\mathrm{eff} \supset \frac{1}{\Lambda^2} \, \sum_f q_f \bar{f} \gamma^\mu f \bar{q}_\mathrm{d} \gamma_\mu q_\mathrm{d} \; ,
\end{align}
with
\begin{align}
\Lambda = \frac{m_{Z'}}{\sqrt{\kappa e e_\mathrm{d}}} \; .
\end{align}
The corresponding production cross section scales like
\begin{align}
\label{eq:belle2_crosssection}
\sigma(f \bar{f} \to q_\mathrm{d} \bar{q}_\mathrm{d}) \propto \frac{s}{\Lambda^4} \; .
\end{align}

\begin{figure}[t]
\begin{center}
\includegraphics[width=\textwidth]{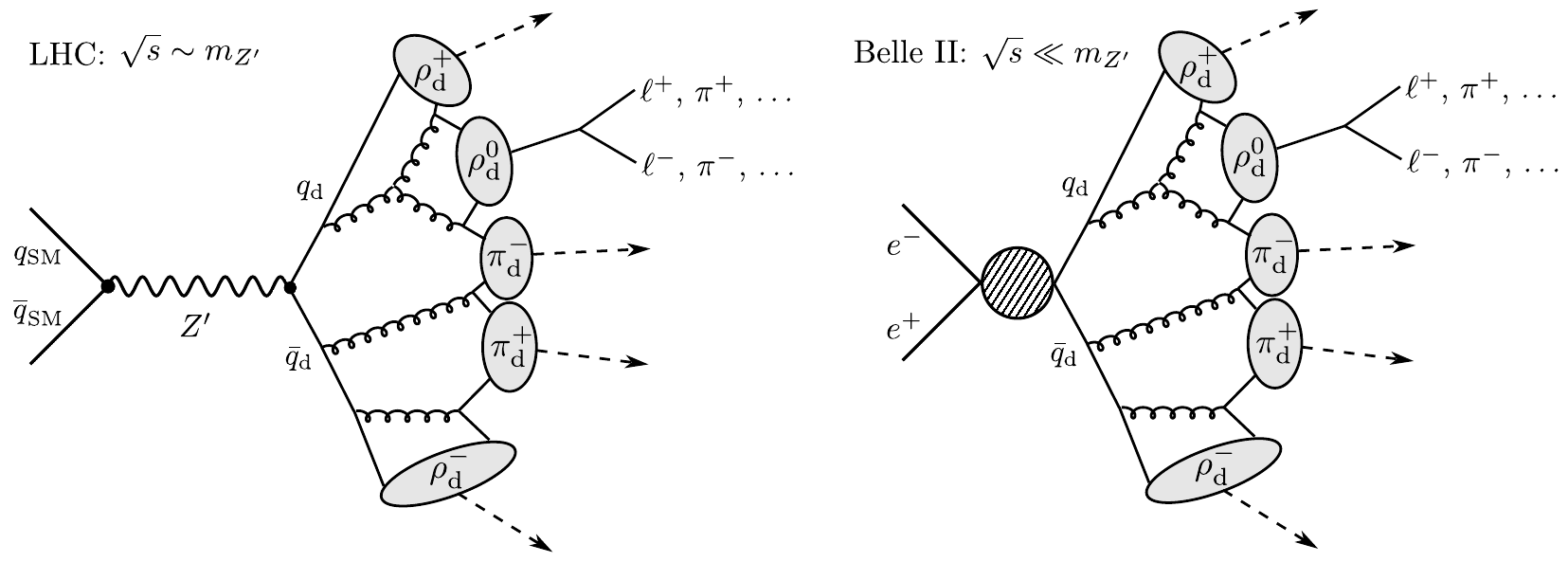}
\caption{Illustration of dark shower production at the LHC (left) via $Z'$ exchange with $\sqrt{s} \sim m_{Z'}$ and at \belletwo (right) via the effective interaction in eq.~(\ref{eq:smfermions_darkquarks_effective}) (represented by the shaded circle) with $\sqrt{s} \ll m_{Z'}$. \label{fig:sketch}}
\end{center}
\vspace{-5mm}
\end{figure}

The dark quarks subsequently undergo fragmentation and hadronisation in the hidden sector resulting in a dark shower consisting of a mix of different dark meson species (see figure~\ref{fig:sketch}). In the model considered here, most of these dark mesons are stable and hence remain invisible to the detector. The only exception are the unstable $\rhodzero$ mesons, which are expected to make up 25~\% of the shower on average\footnote{This estimate is derived from counting the degrees of freedom of approximately mass-degenerate dark pseudoscalar and vector mesons that can be produced in the shower. See Ref.~\cite{Bernreuther:2019pfb} for further details.} and can give rise to visible tracks by decaying to charged leptons or hadrons inside the detector via mixing with the photon-like $Z'$.

For $m_\rhod \ll m_{Z'}$ we can use eq.~(\ref{eq:smfermions_darkquarks_effective}) to obtain the effective interaction of the $\rhodzero$ mesons with SM fermions~\cite{Bernreuther:2019pfb}:
\begin{equation}
\mathcal{L}_\mathrm{eff} \supset \frac{2}{g} \frac{\mrhod^2}{\Lambda^2} \rhodzero^\mu \sum_f q_f \bar{f} \gamma_\mu f \; ,
\end{equation}
where $g$ is the dark pion-vector-meson coupling strength, which we set to $g = 1$ in the following. The partial width for the mixing-induced decay of $\rhodzero$ mesons to pairs of leptons is given by
\begin{align}
\label{eq:gamma_rho0_ll_darkphoton}
\Gamma\left(\rho^0_\mathrm{d}\to \ell^+ \ell^-\right) =
\frac{1}{3\pi g^2}\frac{\mrhod^5}{\Lambda^4}\left(1-4\frac{m_\ell^2}{\mrhod^2}\right)^{1/2} \left(1+2\frac{m_\ell^2}{\mrhod^2}\right) \; .
\end{align}
Since the $\rhodzero$ meson couples to SM fermions in exactly the same way as the SM photon, the hadronic decay width can be written as
\begin{align}
\label{eq:rho0_hadronicwidth}
\Gamma(\rhodzero \to \text{hadrons}) = R(\sqrt{s} = \mrhod) \; \Gamma(\rhodzero \to \mu^+\mu^-) \; ,
\end{align}
where $R(\sqrt{s})$ denotes the cross section ratio $\sigma(e^+e^- \to \text{hadrons})/\sigma(e^+e^- \to \mu^+\mu^-)$ through an off-shell photon at centre-of-mass energy $\sqrt{s}$, which is enhanced in the vicinity of relevant QCD resonances~\cite{Ezhela:2003pp, Zyla:2020zbs}. Analogously, the partial width for a specific hadronic decay mode can be calculated by replacing the inclusive cross section $\sigma(e^+e^- \to \text{hadrons})$ by the production cross section for the appropriate final state, e.g. $\sigma(e^+e^- \to \pi^+\pi^-)$ for the decay $\rhodzero \to \pi^+ \pi^-$.

\begin{figure}[t]
\begin{center}
\includegraphics[width=0.495\textwidth]{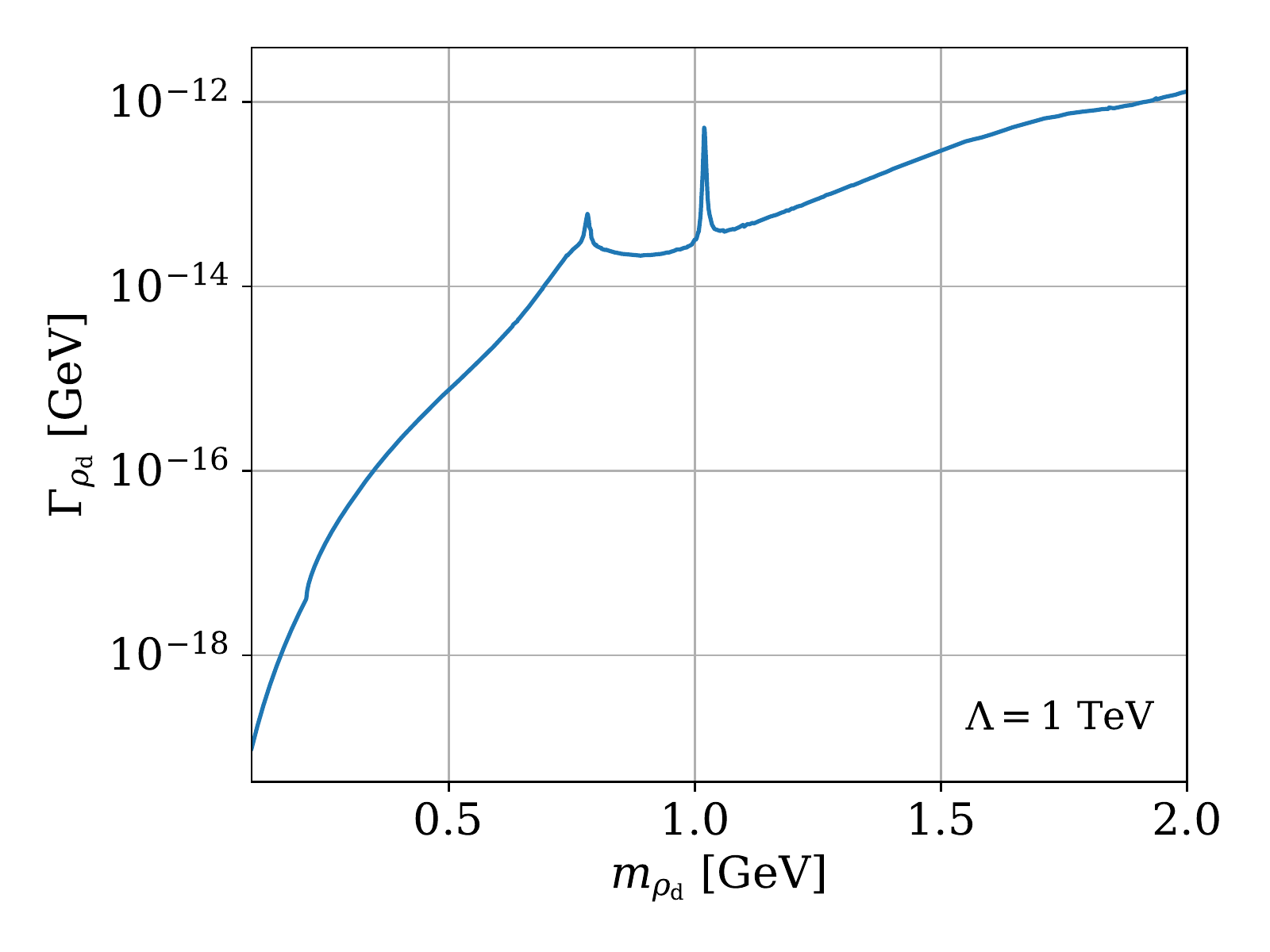}
\includegraphics[width=0.495\textwidth]{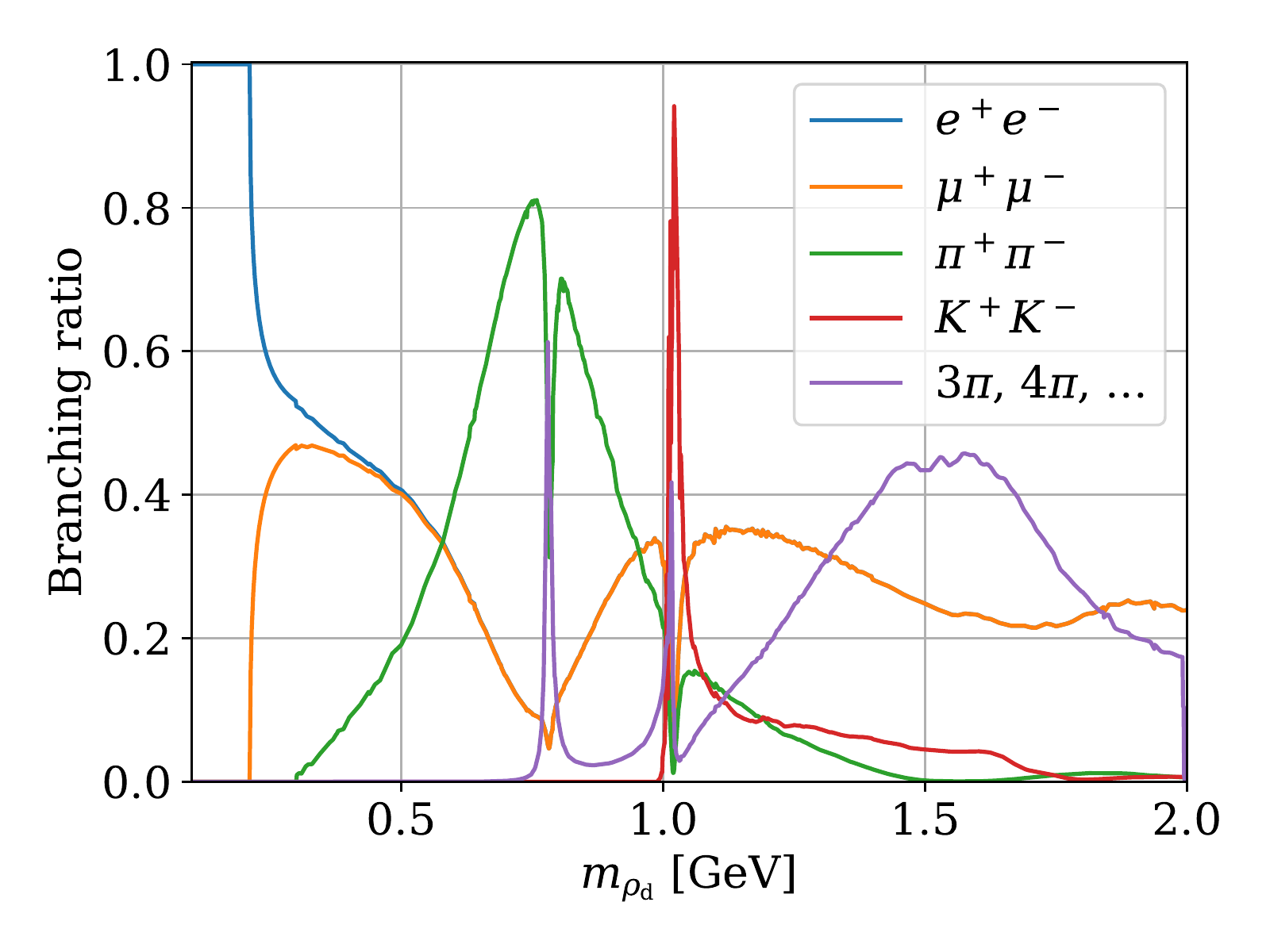}
\caption{
Total $\rhodzero$ meson width (left) and the individual branching ratios (right) as a function of the $\rhodzero$ meson mass $m_\rhod$. \label{fig:BRs}}
\end{center}
\vspace{-5mm}
\end{figure}

If $\mrhod > 2\mpid$, the $\rhodzero$ meson can also decay to pairs of dark pions, in which case the dark sector does not lead to visible collider signatures. We will instead consider the scenario where the $\rhodzero$ mesons and dark pions are sufficiently close in mass for dark pion DM to freeze out via forbidden annihilations into dark vector mesons. This assumption stands in contrast with the expectation that dark pions would be light pseudo-Goldstone bosons in a dark sector with dark quark masses that are negligible compared to the dark confinement scale. However, for dark quark masses not too far below the confinement scale it is plausible that $2\mpid > \mrhod$~\cite{DeGrand:2019vbx}. In this case, the $\rhodzero$ cannot decay into dark sector particles and its total width
\begin{align}
\label{eq:rho0_totalwidth}
\Gamma_{\rhodzero} = \Gamma(\rhodzero \to \text{hadrons}) + \sum_\ell \Gamma(\rhodzero \to \ell^+ \ell^-) \; .
\end{align}
is given by the sum over its SM decay channels.\footnote{For a general strongly interacting sector, three-body decays of the form $\rho_\mathrm{d} \to \pi_\mathrm{d} Z'^* \to \pi_\mathrm{d} f \bar{f}$ are also possible. However, the matrix elements for these decay modes vanish in our model since they are proportional to $\mathrm{Tr}(\{T^a, T^b\}Q)$, with $T^a$ and $T^b$ denoting the flavour $SU(2)$ generators corresponding to the involved $\rho_\mathrm{d}$ and $\pi_\mathrm{d}$, respectively, and $Q = \mathrm{diag}(1,-1)$ being the dark quark charge matrix.} For a given mass, the proper decay length therefore simply scales as $c\tau_{\rho_\mathrm{d}} \propto \Lambda^4$. For instance, for a mass of $\mrhod=500$~MeV, we find
\begin{align}
c\tau_{\rhodzero} = 26~\mathrm{cm} \times \left(\frac{\Lambda}{10^3~\mathrm{GeV}}\right)^4 \; .
\end{align}
The total $\rhodzero$ meson width as a function of $m_\rhod$, as well as the corresponding branching ratios, are shown in figure~\ref{fig:BRs}.

Since the $\rhodzero$ meson decay and dark shower production both depend on the mass and couplings of the mediator only through the effective scale $\Lambda$, there is a direct correspondence between the LLP decay length and the production cross section at $B$ factories for a given LLP mass. Specifically, by combining eqs.~\eqref{eq:belle2_crosssection}, \eqref{eq:gamma_rho0_ll_darkphoton} and \eqref{eq:rho0_totalwidth}, we find that
\begin{align}
\sigma(e^+e^- \to q_\mathrm{d} \bar{q}_\mathrm{d}) \propto \frac{s}{\tau_{\rhodzero}\mrhod^5} \; .
\end{align}
In this low-energy regime, the parameter space of our model hence simplifies tremendously, which allows searches at BaBar and \belletwo to set constraints on the LLP mass and decay length with no dependence on other free parameters.

\subsection{Dark shower simulation}

The primary focus of this work will be on the sensitivity to dark showers of the \belletwo experiment at the SuperKEKB accelerator.
SuperKEKB is a circular asymmetric $e^+e^-$ collider with a nominal collision energy of $\sqrt{s} = 10.58\,\mathrm{GeV}$~\cite{Belle-II:2010dht}.
We generate parton-level events for dark quark pair production $e^+e^- \to q_\mathrm{d} \bar{q}_\mathrm{d}$ with \texttt{MadGraph5\_aMC{@}NLO}~2.6.4~\cite{Alwall:2014hca} using a UFO file for our model created with \texttt{FeynRules}~\cite{Alloul:2013bka}. 
Shower and hadronisation in the dark sector are simulated with the Hidden Valley module\footnote{Note that this module requires the dark hadronisation scale to be above approximately 100 MeV, which is why we constrain ourselves to $\rhod$ meson masses $\mrhod \ge 100\,\mathrm{MeV}$ in the following.} of \texttt{Pythia~8}~\cite{Sjostrand:2014zea, Carloni:2010tw, Carloni:2011kk}. While the shower is under perturbative control, hadronisation in a strongly interacting hidden sector is associated with substantial uncertainty. The Hidden Valley module of \texttt{Pythia~8} employs the same Lund string fragmentation model that is also used for QCD. The dimensionless parameters of the string model are chosen identical to the QCD sector, while the parameters with non-zero mass dimension are scaled by the appropriate power of the ratio of the dark confinement scale to the QCD confinement scale. While the values of these parameters in a realistic dark sector are uncertain, we emphasize that the model-agnostic analyses discussed in section~\ref{sec:DV} do not depend on details of the dark sector hadronisation. After hadronisation, each $\rhodzero$ meson produced in the dark shower is decayed to leptons or QCD mesons with branching fractions following from the partial widths given in eqs.~\eqref{eq:gamma_rho0_ll_darkphoton} and \eqref{eq:rho0_hadronicwidth}.

\begin{figure}[t]
\begin{center}
\includegraphics[width=0.495\textwidth]{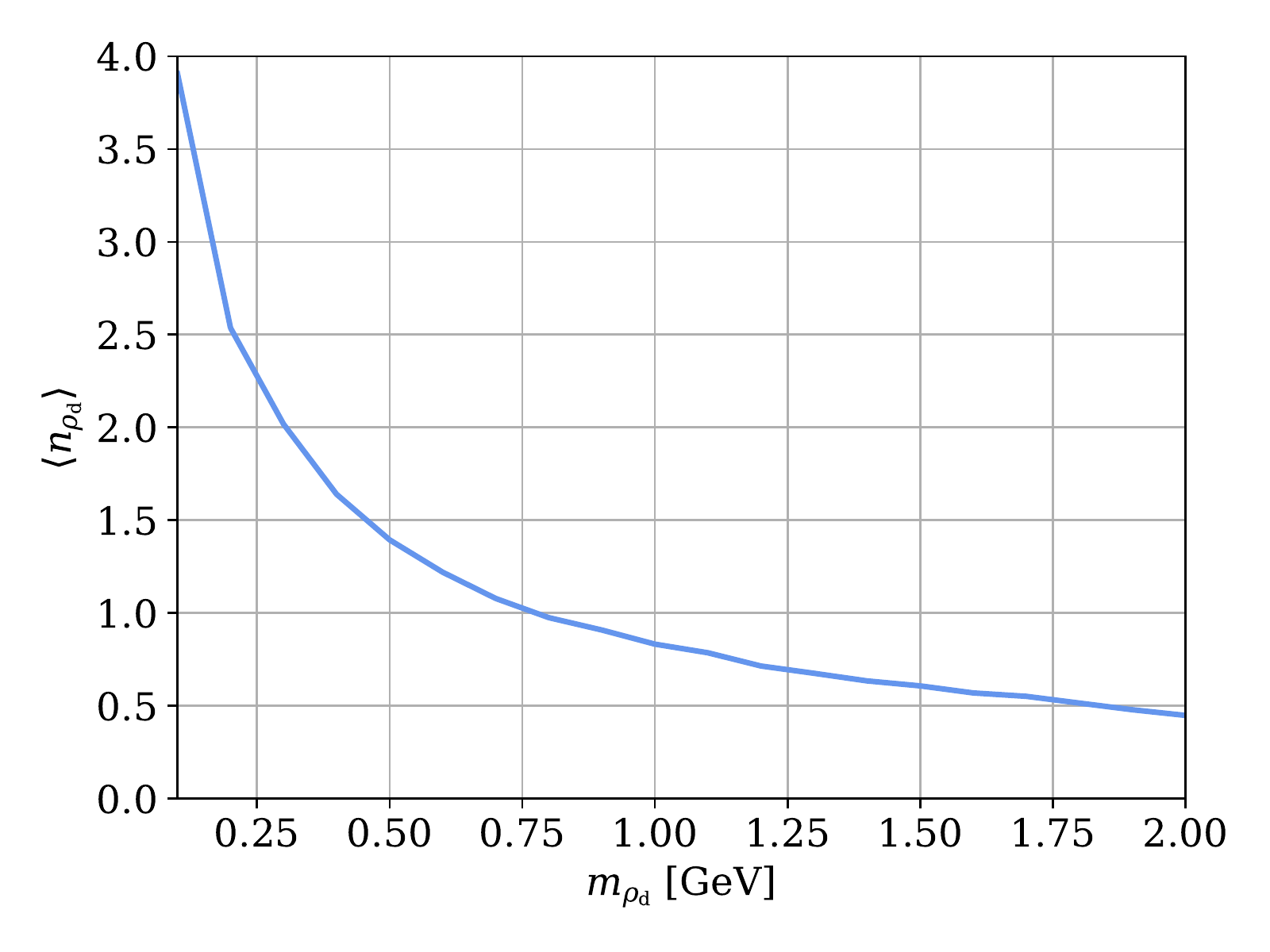}
\includegraphics[width=0.495\textwidth]{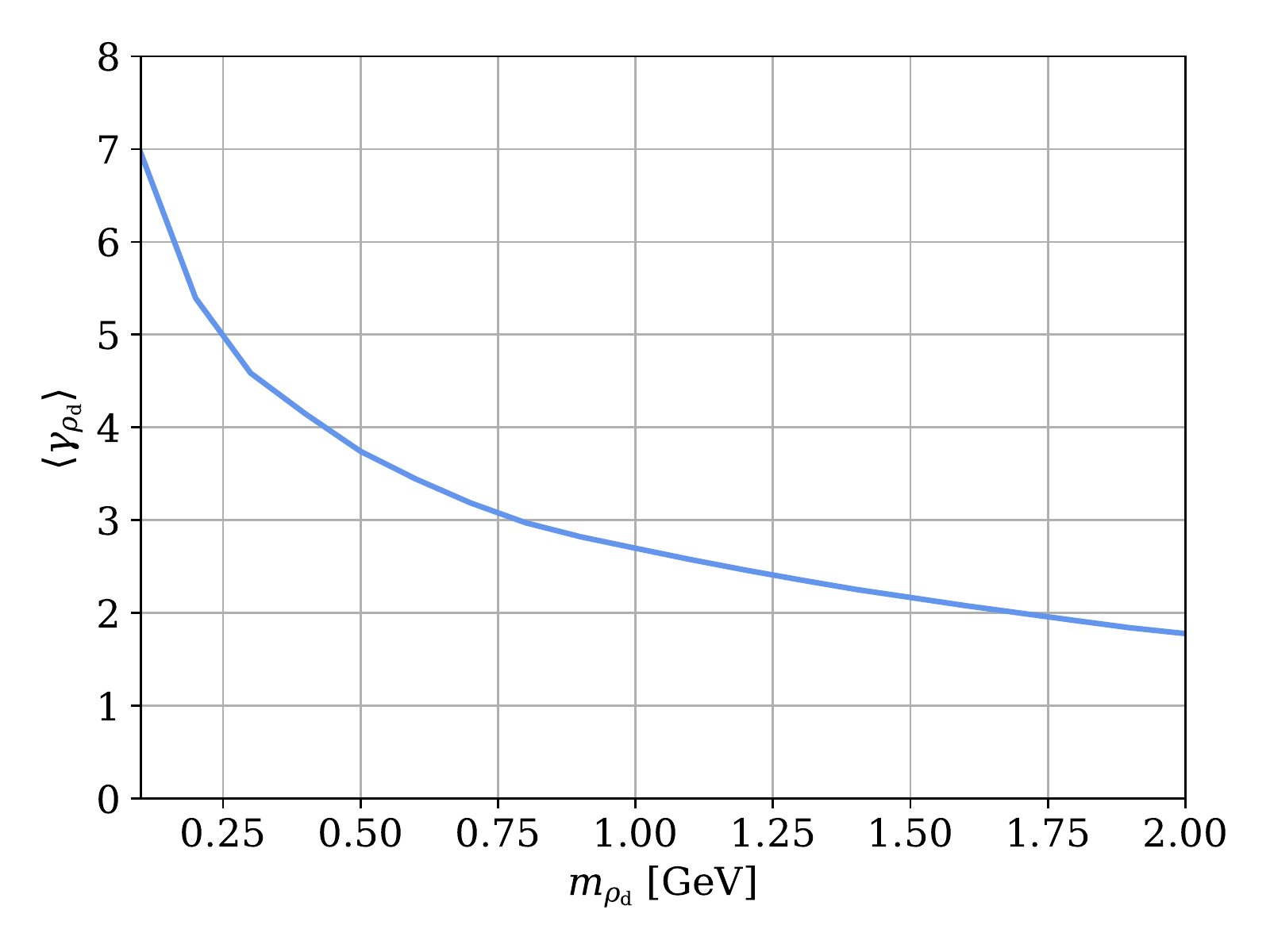}
\caption{\label{fig:belle2_llp_mult_boost}Average multiplicity (left) and average boost in the laboratory frame (right) of $\rhodzero$ mesons in dark showers produced at \belletwo ($\sqrt{s} = 10.58\,\mathrm{GeV}$) as a function of $m_{\rho_\mathrm{d}}$.}
\end{center}
\vspace{-5mm}
\end{figure}

 Since 25~\% of all dark mesons in the shower are expected to be $\rhodzero$ mesons and each $\rhodzero$ decays visibly, the average invisible fraction of the dark shower is $r_\mathrm{inv}=0.75$ as long as all decays happen inside the fiducial detector volume. However, it is one of the most salient characteristics of the dark shower signal that the fraction of visibly decaying dark mesons varies substantially around this average from event to event. Moreover, the absolute number of $\rhodzero$ mesons in each dark shower depends on their mass. Figure~\ref{fig:belle2_llp_mult_boost} shows the average number of $\rhodzero$ LLPs per event and their boost as a function of $\mrhod$. Both multiplicity and boost decrease with increasing LLP mass. Nonetheless, in the sub-GeV mass range many events contain one or even multiple $\rhodzero$ mesons, which suggests that LLP searches at $B$ factories are sensitive to dark showers consisting of light dark mesons. We discuss these searches in the following section. In addition, we can infer from figure~\ref{fig:belle2_llp_mult_boost} that for LLP masses $\mrhod \gtrsim 500$~MeV a significant fraction of events contains no $\rhodzero$ mesons and hence remains invisible. If such an invisible dark shower is produced in association with an ISR photon, it can give rise to a single-photon signal, which we consider in detail in section~\ref{sec:single_photon}.

\section{Sensitivity of displaced vertex searches at $B$ factories}
\label{sec:DV}

Having characterised the properties of dark shower production through an effective portal  in our model, we next study the sensitivity to this signal of searches for LLPs at $B$ factories. In this section we briefly summarise the features of an existing model-independent search for LLPs at BaBar and of a proposed search for LLPs at \belletwo.

\subsection{Model-independent search for LLPs at BaBar}

The BaBar collaboration has carried out a model-independent search for LLPs decaying to two oppositely charged tracks with an integrated luminosity of $489.1$~fb$^{-1}$~\cite{BaBar:2015jvu}. The relevant LLP decay modes are $e^+e^-$, $\mu^+\mu^-$, $e^\pm\mu^\mp$, $\pi^+\pi^-$, $K^+K^-$ and $K^\pm\pi^\mp$.  Displaced vertices are reconstructed from tracks originating from decays at a transverse distance $R$ between $1$~cm and $50$~cm from the interaction point. Moreover, each track has to fulfil the requirement that $d_0/\sigma_{d_0} > 3$, where $d_0$ denotes the transverse impact parameter of the track and $\sigma_{d_0}$ its experimental uncertainty. 
Depending on the particle species, the invariant mass of the LLP decay products has to fulfil $m_{e^+e^-} > 0.44$~GeV, $m_{\pi^+\pi^-} > 0.86$~GeV, $m_{K^+K^-} > 1.35$~GeV, or either $m_{\mu^+\mu^-} < 0.37$~GeV or $m_{\mu^+\mu^-} > 0.5$~GeV. To ensure that our signal events pass the Level 1 trigger, we require at least three charged tracks with transverse momentum $p_T > 0.12$~GeV, at least two charged tracks with $p_T > 0.18$~GeV, and at least two clusters with energy $E > 0.13$~GeV~\cite{BaBar:2001yhh}. For our model, this essentially means that only dark shower events that contain at least two $\rho_\mathrm{d}^0$ LLPs are accepted.

By searching for a peak in the distribution of the two-track invariant mass above a smooth background, the analysis places a model-independent limit on the fiducial cross section
\begin{align}
\label{eq:babar_limit}
\sigma_\text{fid} = \sigma(e^+e^- \to L X) \; \mathrm{BR}(L \to \mathcal{F}) \; \epsilon(\mathcal{F}) \; ,
\end{align}
with the production cross section $\sigma(e^+e^- \to L X)$ of the LLP $L$, its branching ratio $\mathrm{BR}(L \to \mathcal{F})$ for a given final state $\mathcal{F}$ and the associated experimental efficiency $\epsilon(\mathcal{F})$. The efficiencies are tabulated as a function of the proper decay length $c\tau$, mass $m$ and truth-level transverse momentum $p_T$ of the LLP in the centre-of-mass system of BaBar. For each final state, the stated efficiency includes the detector acceptance, trigger, reconstruction and selection efficiency, which allows us to reinterpret the reported bound in terms of long-lived $\rhodzero$ mesons in our dark shower model. 
We note that the efficiencies necessary for recasting are available only for $m > 500\,\mathrm{MeV}$ and $c\tau > 5$~mm, so that no BaBar exclusion limits on dark showers can be determined for smaller masses or decay lengths.

\subsection{Sensitivity of an LLP search at \belletwo}

\begin{table}[t!]
    \caption{\label{tab:belle2_trigger} Overview of the various trigger conditions sensitive to two-particle final states obtained by combining \belletwo hardware and software triggers.}
    \begin{center}
      \begin{tabular}{l p{7cm} p{4cm}}
        \toprule
        \textbf{Name} & \textbf{Condition} & \textbf{Note} \\
        \midrule
\emph{2 GeV energy} & At least one calorimeter cluster with $E_\text{CMS} > \unit[2]{GeV}$ and {$22^\circ < \theta_\text{lab} < 139.3^\circ$} & Not efficient for muons  and  hadrons\\
\midrule
\emph{Two tracks} & Two tracks with \mbox{$38^\circ < \theta_\text{lab} < 127^\circ$} and a transverse momentum $p_T > \unit[300]{MeV}$ each, as well as an azimuthal opening angle at the interaction point in the lab system $ \Delta \varphi > 90^\circ$ & Not efficient beyond a radius of $R_\text{max} = \unit[17]{cm}$\\
\midrule
\emph{One muon} & One track with \mbox{$38^\circ < \theta_\text{lab} < 127^\circ$} and a transverse momentum $p_T > \unit[900]{MeV}$ & Only efficient for muons and hadrons; not efficient beyond a radius of $R_\text{max} = \unit[17]{cm}$\\
\midrule
\emph{1 GeV $E$ sum} & The sum of all clusters with \mbox{$E_\text{lab} > \unit[100]{MeV}$} and $27^\circ < \theta_\text{lab} < 128^\circ$ is larger than 1 GeV & Muons and hadrons are assumed to contribute at most $200 \, \mathrm{MeV}$\\
\midrule
\emph{Displaced vertex} & At least one displaced vertex in the event with $\unit[0.9]{cm} < R < \unit[60]{cm}$, formed from two tracks with $p_T > \unit[100]{MeV}$ and \mbox{$38^\circ < \theta_\text{lab} < 127^\circ$} each & Not yet implemented\\
  \bottomrule
      \end{tabular}
  \end{center}
\end{table}

\begin{table}[t!]
    \caption{\label{tab:belle2_cuts} Selection requirements after triggers for the model-independent search for LLPs at \belletwo proposed in Refs.~\cite{Duerr:2019dmv, Duerr:2020muu}. Note that the selection requirements for 
    charged pions and kaons are identical to those for muons.
    Here $\alpha$ denotes the opening angle of the pair of particles and $m$ their invariant mass. $R$ and $z$ are the distances of the decay vertex from the interaction point in transverse and longitudinal direction, respectively. The angle $\theta_\text{lab}$ is the angle between the position vector of the decay vertex and the $z$-axis in the laboratory frame.}
    \begin{center}
      \begin{tabular}{cc}
        \toprule
        electron pairs &  muon pairs \\
        \midrule
        $p(e^+), \, p(e^-) > 0.1$~GeV & $p(\mu^+), \, p(\mu^-) > 0.05$~GeV \\
        $m_{e^+e^-} > 0.03$~GeV & $m_{\mu^+\mu^-} < 0.48$~GeV \emph{or} $m_{\mu^+\mu^-} > 0.52$~GeV \\
        $\alpha(e^+,e^-) > 0.025$~rad & \\
        \midrule
        \multicolumn{2}{c}{Displaced vertex position}\\
        \midrule
        $0.2~\text{cm} < R < 0.9~\text{cm}$ \emph{or} $17~\text{cm} < R < 60~\text{cm}$ & $0.2~\text{cm} < R < 60~\text{cm}$ \\
        \multicolumn{2}{c}{$-55~\text{cm} \leq z \leq 140~\text{cm}$} \\
        \multicolumn{2}{c}{$17^\circ \leq \theta_\text{lab} \leq 150^\circ$} \\
        \bottomrule
      \end{tabular}
  \end{center}
\end{table}

To predict the sensitivity of \belletwo to dark showers we consider the search for LLPs proposed in Refs.~\cite{Duerr:2019dmv, Duerr:2020muu}, which searches for displaced decays of LLPs into pairs of oppositely charged leptons, pions or kaons. We consider events that pass at least one of the existing \belletwo triggers or the planned displaced vertex trigger, see table~\ref{tab:belle2_trigger} and Ref.~\cite{Duerr:2020muu}. After passing the trigger, events are selected if they contain at least one displaced LLP decay into the relevant final states and fulfils the additional criteria summarised in table~\ref{tab:belle2_cuts}. For displaced $\mu^+\mu^-$ pairs, backgrounds are expected to be suppressed to a negligible level for the integrated luminosity we consider. For displaced $e^+e^-$ pairs, we still expect a sizeable photon conversion background in the vertex detectors, primarily from radiative Bhabha scattering with both tracks out of the detector acceptance. Therefore, we do not consider LLP decays to $e^+e^-$ at transverse distances $0.9~\text{cm} < R < 17~\text{cm}$ in the analysis. For more detailed discussions about possible remaining backgrounds see Ref.~\cite{Ferber:2022ewf}.

We find that different triggers are necessary to target the different possible final states. If the LLP decays into an electron-positron pair, the most sensitive trigger is found to be the \emph{1 GeV $E$ sum} trigger. This trigger is however inefficient for other final states, since we assume that each muon and hadron deposits at most 200 MeV in the calorimeter (and the contribution from events with more than one LLP is small). Hence, for muons and hadrons, the most efficient triggers turn out to be the \emph{two tracks} trigger for small $\rhodzero$ meson masses and the \emph{one muon} trigger for large $\rhodzero$ meson masses. By assumption, these triggers are not efficient for $R > 17 \, \mathrm{cm}$. Hence, a dedicated displaced vertex trigger offers gains in sensitivity towards large decay lengths (i.e.\ small couplings). 

\begin{figure}[t]
\begin{center}
\includegraphics[width=0.495\textwidth]{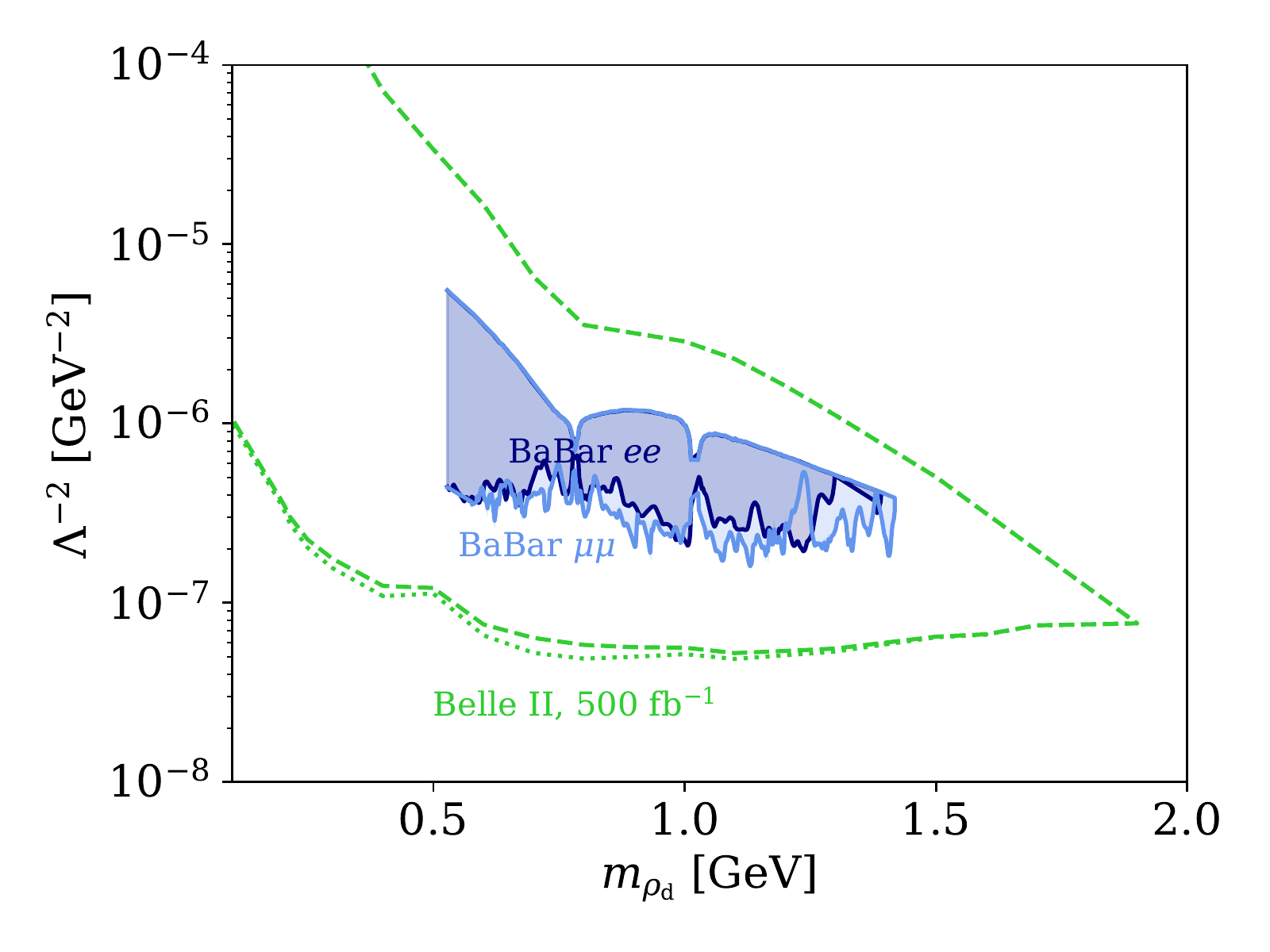}
\includegraphics[width=0.495\textwidth]{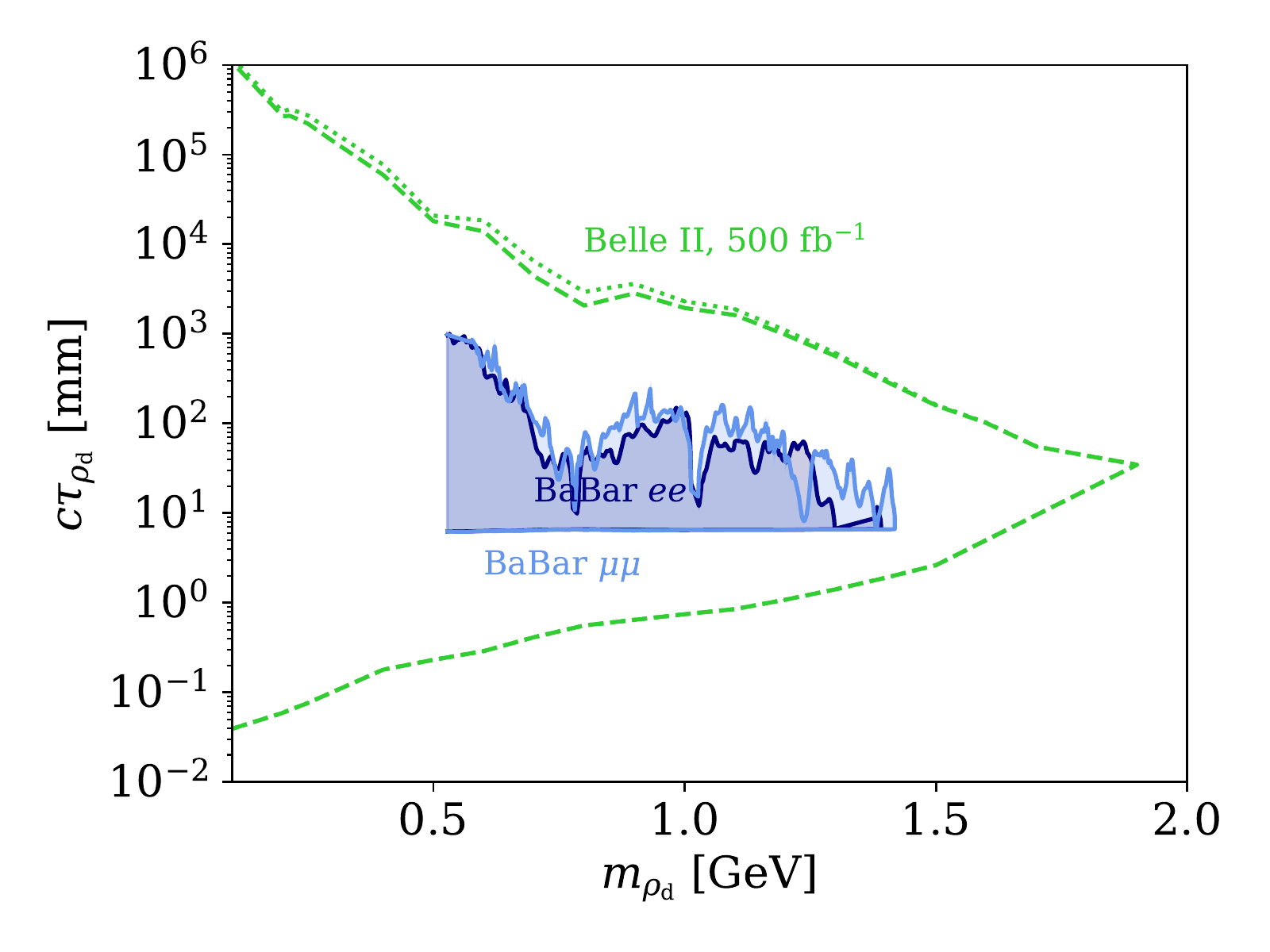}
\caption{Projected sensitivity to the effective production of dark showers of a displaced vertex search at \belletwo compared to existing exclusion limits from BaBar in terms of the effective coupling $\Lambda^{-2}$ (left) and the $\rhodzero$ meson decay length $c\tau_{\rhod}$ (right). We show separately the sensitivity when using existing triggers only (dashed) and when furthermore employing a displaced vertex trigger (dotted).
\label{fig:BaBarBelle}
}
\end{center}
\vspace{-5mm}
\end{figure}

In figure~\ref{fig:BaBarBelle} we compare the 90\% confidence level exclusion limits obtained from our reinterpretation of the BaBar analysis with the sensitivity projections for \belletwo based on an integrated luminosity of $500\,\mathrm{fb}^{-1}$, which is comparable to the amount of data to be collected by summer 2022. We show separately the projections obtained with and without DV trigger to highlight the benefits of such a trigger for extending the sensitivity towards long lifetimes. Since the proper decay length $c\tau_\rhod$ is a function of $\Lambda$ and $m_\rhod$ only (for the assumed coupling structure), the two panels of figure~\ref{fig:BaBarBelle} are fully equivalent and no additional parameters need to be specified. This highlights the benefits of the effective description of dark shower production introduced in section~\ref{sec:eff}.

\begin{figure}[t]
\begin{center}
\includegraphics[width=0.495\textwidth]{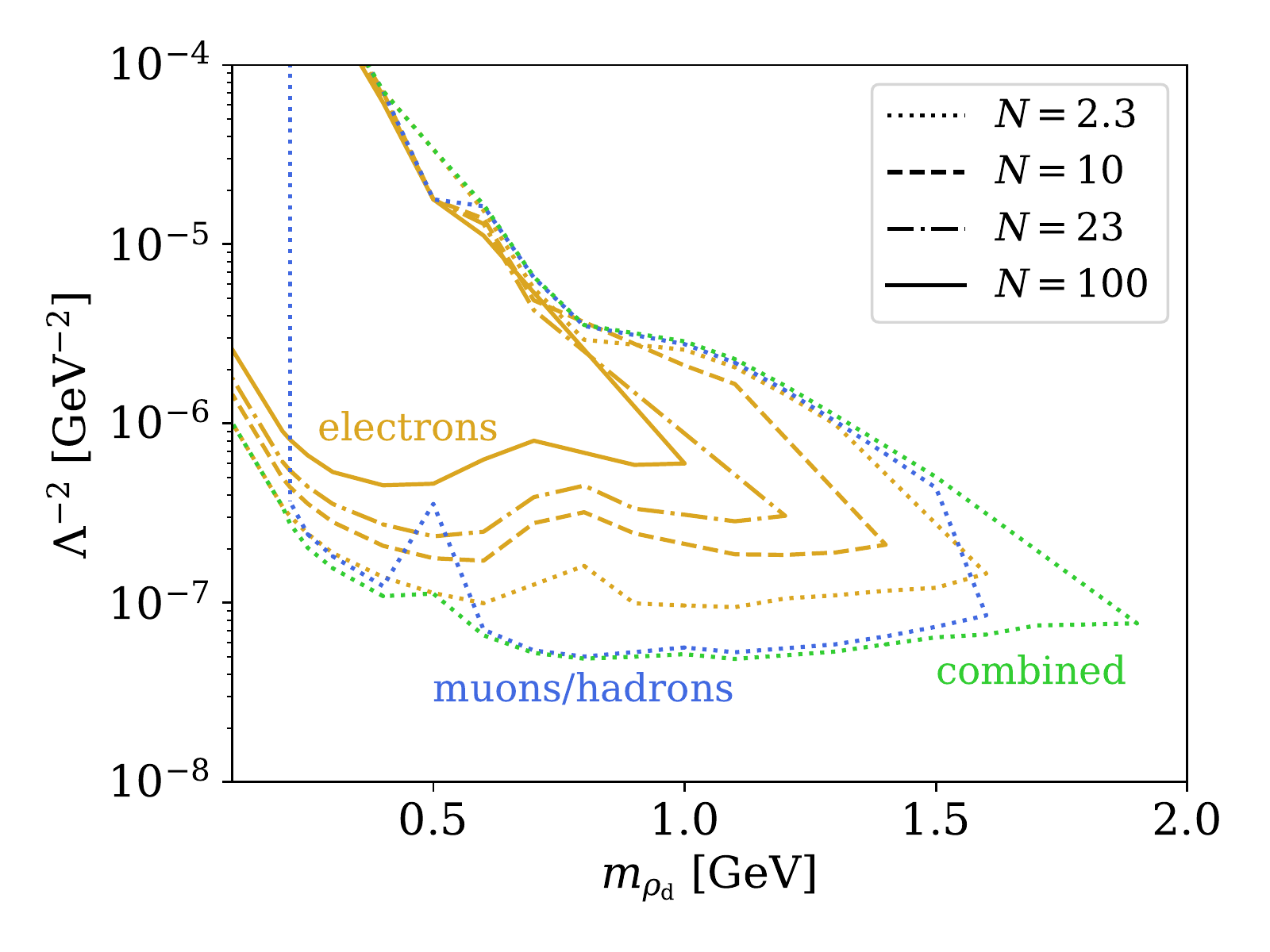}
\includegraphics[width=0.495\textwidth]{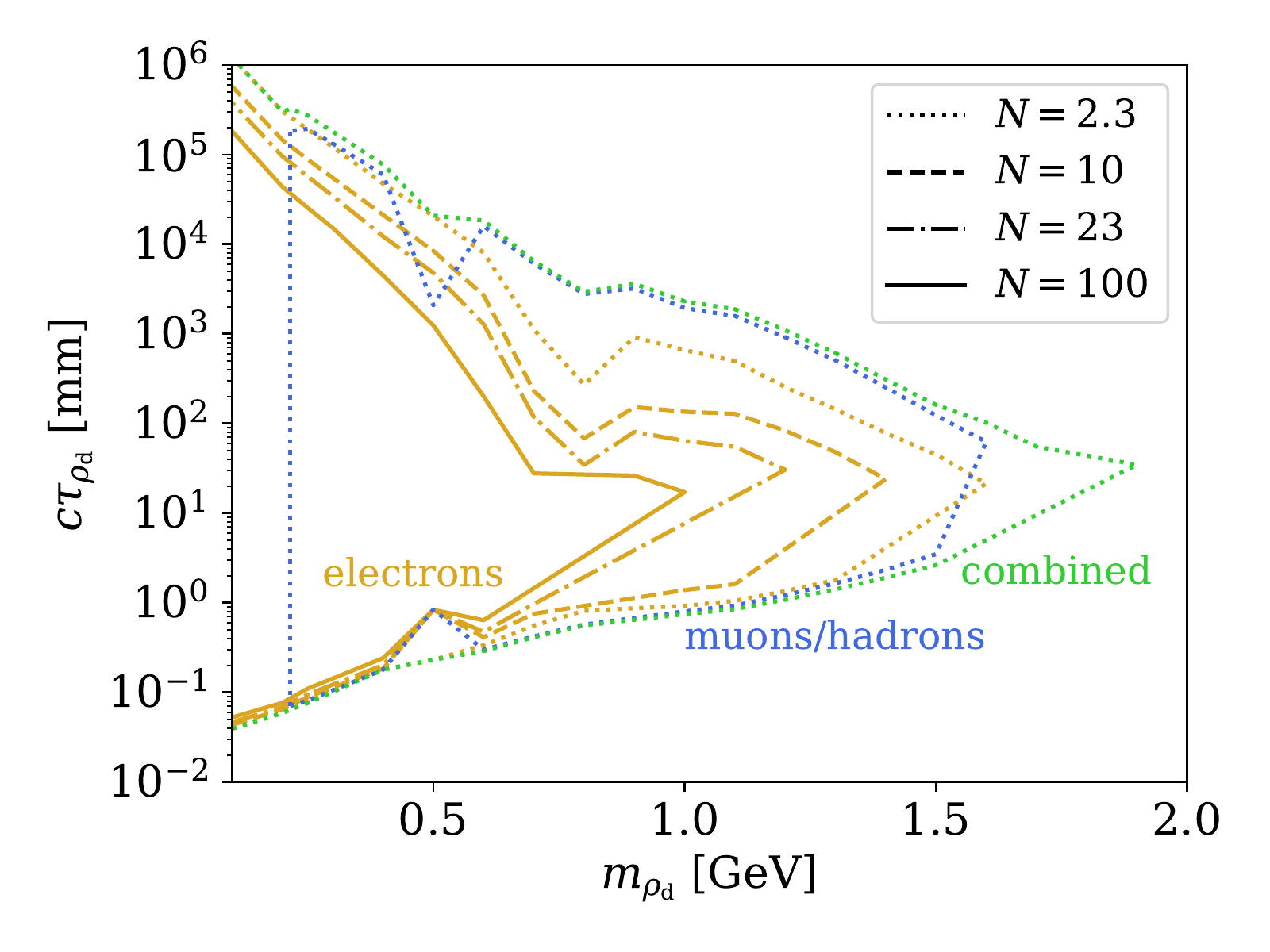}
\caption{Same as figure~\ref{fig:BaBarBelle}, but showing separately the sensitivity for the electron channel and the muon/hadron channel in \belletwo. For the former, we show different contours corresponding to different numbers of predicted events in order to illustrate the degradation of the sensitivity in case of non-negligible backgrounds.
\label{fig:backgrounds}
}
\end{center}
\vspace{-5mm}
\end{figure}

To conclude this section, let us briefly return to the question of experimental backgrounds in the electron channel. If these turn out to be non-negligible and impossible to remove through tighter cuts on the invariant mass or the opening angle, it may be advantageous to perform separate analyses for different final states. In figure~\ref{fig:backgrounds} we show the sensitivity that can be achieved by analysing separately the electron channel and the muon/hadron channel. For the former, we show four different contours corresponding to 2.3, 10, 23 and 100 predicted events in order to illustrate how the sensitivity would degrade in the presence of backgrounds. We find that, in the absence of backgrounds, the sensitivity of both channels is comparable above the muon threshold, such that the muon/hadron channel dominates the sensitivity as soon as there are relevant backgrounds in the electron channel.\footnote{Note that in our analysis, the sensitivity of the muon/hadron channel does not vanish for $m_\rhod = 500\,\mathrm{MeV}$ even though we exclude events with $480\,\mathrm{MeV} < m_{\mu^+\mu^-} < 520\,\mathrm{MeV}$. The reason is that our \texttt{Pythia} simulations yield a small fraction of events with $m_{\mu^+\mu^-} < m_\rhod$ because of final-state radiation. However, these events may be difficult to distinguish from background in practice, such that our sensitivity estimates for $m_\rhod = 500\,\mathrm{MeV}$ are likely too optimistic.}

For the remainder of this paper, we will assume that backgrounds in the electron channel can be suppressed to a negligible level and that a dedicated DV trigger will indeed be implemented. Hence, we will show the most optimistic projection (i.e. the green dotted line in figures~\ref{fig:BaBarBelle} and~\ref{fig:backgrounds}) from now on.

\section{Comparison of constraints on the low-energy effective theory}
\label{sec:lowe}

In this section we compare the sensitivity projections for \belletwo derived above to other constraints and projections from low-energy experiments, which can be calculated using the effective description derived in section~\ref{sec:eff}. Constraints from high-energy experiments, for which the effective description becomes invalid, will be considered in section~\ref{sec:highe}.

\subsection{Single-photon searches}
\label{sec:single_photon}

The model that we consider can also give rise to the single-photon signature ($e^+ e^- \to \gamma + \text{invisible}$) if a pair of dark quarks is produced together with a photon from initial state radiation and then remains fully invisible. This can happen if either only stable dark sector states are produced in the hadronisation process or if all unstable states produced decay outside the detector acceptance.

A crucial difference to typical single-photon searches is however that the dark quark pair is produced via a higher-dimensional operator and therefore does not have a fixed invariant mass $M \equiv m_{q_\mathrm{d}\bar{q}_\mathrm{d}}$. Correspondingly, the photon energy in the centre-of-mass frame, given by
\begin{equation}
 E_\gamma = \frac{1}{2}\left(\sqrt{s} - \frac{M^2}{\sqrt{s}}\right) \; ,
\end{equation}
is distributed continuously, making the signal much harder to distinguish from potential backgrounds. Indeed, there has been so far no dedicated experimental search for this signature, and the only available data set that could be used for reinterpretation stems from an unpublished BaBar analysis~\cite{BaBar:2008aby} (considered in Ref.~\cite{Essig:2013vha}). Rather than attempting to interpret this data set, we will here focus on the potential sensitivity of \belletwo to the single-photon signature.

The fact that the invariant mass $M$ (i.e.\ the momentum transferred to the dark quark pair) varies from event to event significantly complicates event generation. For $M \gg \Lambda_\mathrm{d}, m_\rhod$ the production of a pair of dark quarks proceeds largely independently from the subsequent hadronisation, such that the corresponding cross section becomes independent of $m_\rhod$.\footnote{We remind the reader that \texttt{Pythia} does not distinguish between the dark pion mass and the $\rhodzero$ meson mass.} For smaller $M$, on the other hand, we need to rely on the differential cross section in $M^2$ provided by \texttt{Pythia}. For $M > 3.5 m_\rhod$ this cross section is obtained using the standard string fragmentation model. For $M < 3.5 m_{\rho_\mathrm{d}}$, on the other hand, \texttt{Pythia} considers a so-called ministring, which yields a pair of dark pions. Since the parton shower is not expected to be valid in this regime, and in order to obtain conservative estimates, we will neglect the region $M < 3.5 m_\rhod$ in our analysis. With these restrictions, we can then obtain the distribution of ISR photons in terms of $E_\gamma$ and the angle $\theta$ in the laboratory frame and determine the probability that the dark quark pair evades detection.

To estimate the sensitivity of \belletwo for this signature we implement the analysis proposed in Ref.~\cite{Belle-II:2018jsg} for low dark photon masses, for which the vast majority of SM background events are removed.\footnote{It is tempting to propose even more stringent cuts, for which even the remaining simulated backgrounds would be removed~\cite{Liang:2021kgw}. However, it seems unlikely that this search can ever be completely background-free, which is why we prefer to keep a non-negligible background in order to obtain more realistic sensitivity estimates.} Specifically, we apply the following cuts:
\begin{equation}
5.399 E_\gamma^2 - 58.82 E_\gamma+ 195.71 < \theta < -7.982 E_\gamma^2 + 87.77 E_\gamma - 120.6
\end{equation}
with $E_\gamma$ in GeV. These cuts can only be satisfied for $E_\gamma > 3.0 \, \mathrm{GeV}$, corresponding to $M^2 < 48.8 \, \mathrm{GeV^2}$. The surviving backgrounds are provided in Ref.~\cite{Belle-II:2018jsg} and can be summarised as follows:
\begin{itemize}
 \item For $M^2 < 10\,\mathrm{GeV^2}$ virtually no background events are found in the simulation.
 \item For $10\,\mathrm{GeV^2} < M^2 < 35 \, \mathrm{GeV^2}$ the background is approximately constant at a level of approximately 125 events/GeV$^2$ for an integrated luminosity of $500\,\mathrm{fb^{-1}}$.
 \item For $35 \, \mathrm{GeV^2} < M^2$ the background rises slowly up to about 375 events/GeV$^2$ for an integrated luminosity of $500\,\mathrm{fb^{-1}}$.
\end{itemize}
Since the background estimate for small $M^2$ is likely too optimistic, we will exclude this range from our analysis and instead focus on the range $10\,\mathrm{GeV^2} < M^2 < 45 \,\mathrm{GeV^2}$, which we divide into 7 equidistant bins.\footnote{The assumed number of background events for an integrated luminosity of $500\,\mathrm{fb^{-1}}$ are then given by [640,  640,  640,  640,  640,  960, 1600].}

We assume that \belletwo will see a number of events compatible with the simulated backgrounds but that the understanding of the backgrounds will be insufficient to allow for background subtraction. Hence, only signals can be excluded that predict significantly more events than observed. Specifically, we treat the background in each bin as a nuisance parameter, such that we obtain the profile log likelihood
\begin{equation}
 -2 \log \mathcal{L} = 2 \sum_i p_i - n_i + n_i \log(n_i / p_i) \; ,
\end{equation}
where $n_i$ is the observed number of events in bin $i$ (assumed to be equal to the simulated background $b_i$) and $p_i$ is the predicted number of events, which is given by $p_i = \text{max}(b_i, s_i)$ with the signal prediction $s_i$. At 90\% confidence level, \belletwo is then expected to be sensitive to parameter points with $-2 \log \mathcal{L} > 2.71$.

\begin{figure}[t]
\begin{center}
\includegraphics[width=0.5\textwidth]{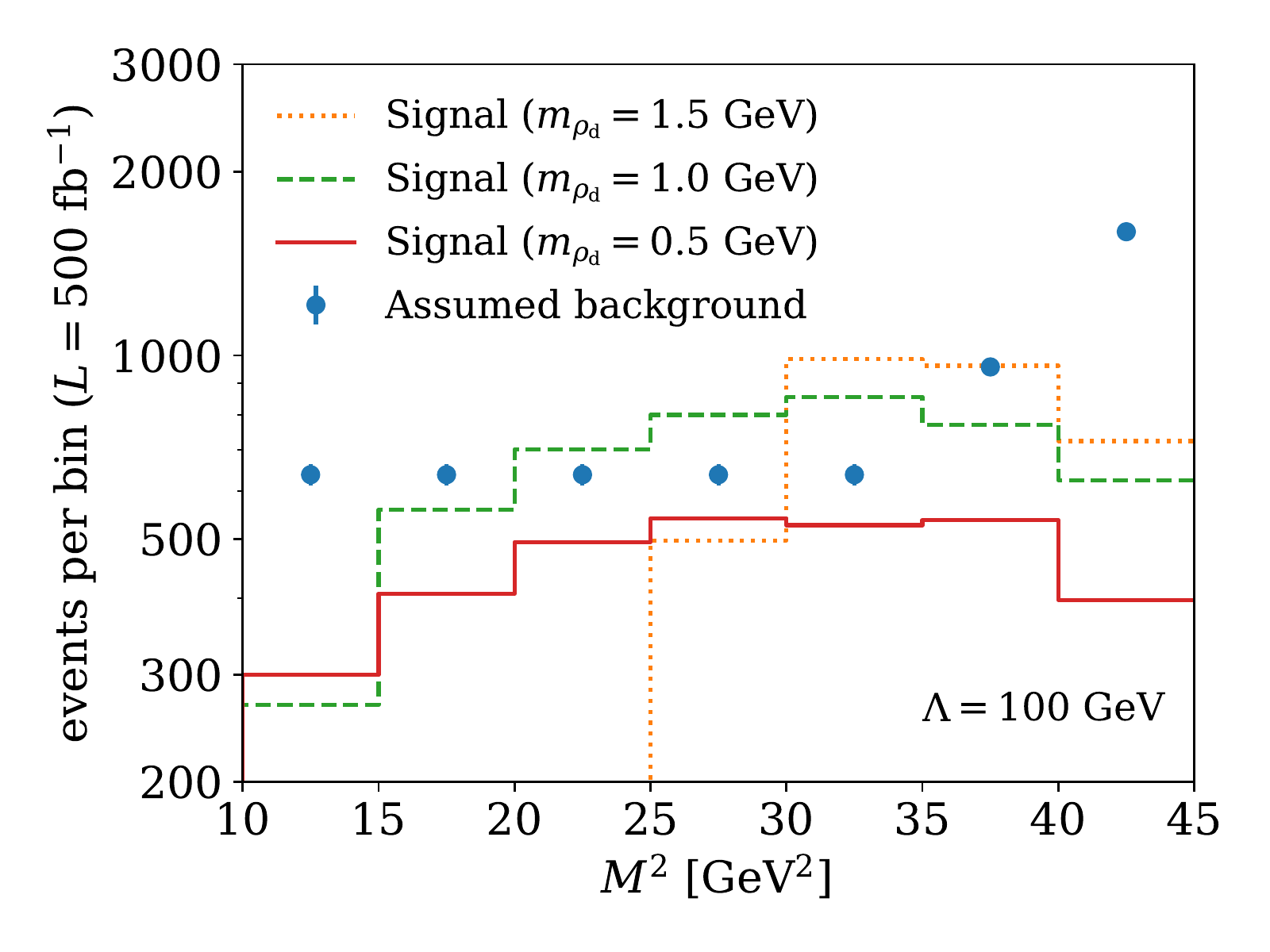}
\caption{Predicted events in the single-photon channel of the \belletwo experiment for the dark shower model with different values of $m_\rhod$ compared to the assumed backgrounds for $500\,\mathrm{fb^{-1}}$. \label{fig:events}}
\end{center}
\vspace{-5mm}
\end{figure}

We find that for small $m_\rhod$ the simulated signals are relatively flat in $M^2$ such that the impact of the chosen binning is expected to be minimal (see figure~\ref{fig:events}).
For larger values of $m_\rhod$, there is a notable feature in the spectrum at $M^2 = 3.5^2 m_\rho^2$ resulting from the cut imposed by our simulation requirements (see above). The sensitivity of the analysis that we consider therefore decreases with increasing $\rhodzero$ meson masses, and it becomes completely insensitive for $m_{\rho_\mathrm{d}} \gtrsim 1.9 \, \mathrm{GeV}$. For small $\rhodzero$ meson masses, on the other hand, the $\rhodzero$ meson multiplicity increases rapidly (see figure~\ref{fig:belle2_llp_mult_boost}) and hence there is a greater chance that the dark shower is vetoed, leading to an overall suppression of the predicted event rate. Hence, for the choice $\Lambda = 100 \, \mathrm{GeV}$ and an integrated luminosity of $500\,\mathrm{fb}^{-1}$, the \belletwo experiment should be able to test the two parameter points with $m_\rhod = 1\,\mathrm{GeV}$ and $m_\rhod = 1.5\,\mathrm{GeV}$, but not the parameter point with $m_\rhod = 0.5\,\mathrm{GeV}$.

\subsection{Resonance searches}

Another way to produce dark sector states at $B$ factories is via the process $e^+ e^- \to \gamma \rho_\mathrm{d}^0$ considered previously in Ref.~\cite{Hochberg:2017khi,Berlin:2018tvf}. Since in our set-up the $\rhodzero$ meson couples exactly like a dark photon $A'$ with kinetic mixing parameter $\kappa$, the corresponding cross section can be obtained from the process $e^+ e^- \to \gamma A'$ through the replacement
\begin{equation}
    e\kappa \to \frac{2}{g} \frac{m_{\rho_\mathrm{d}}^2}{\Lambda^2} \; .
\end{equation}
This relation can be used to directly translate published bounds on $\kappa$ into bounds on $\Lambda^{-2}$.

In the parameter region of interest, the $\rhodzero$ meson is predicted to decay promptly.\footnote{For a detailed discussion of the sensitivity of \belletwo to long-lived vector bosons, we refer to Refs.~\cite{Ferber:2022ewf,Bandyopadhyay:2022klg}.} The most interesting constraints in the present context are therefore the ones searching for a narrow resonance in the $e^+e^-$, $\mu^+ \mu^-$ or $\pi^+ \pi^-$ final state. Specifically, we consider bounds from KLOE~\cite{KLOE-2:2016ydq,KLOE-2:2018kqf} and BaBar~\cite{BaBar:2014zli}, which we obtain using \texttt{Darkcast}~\cite{Ilten:2018crw}. We note that similar constraints are expected from future searches for visibly decaying dark photons at \belletwo.

\bigskip

\begin{figure}[t]
\begin{center}
\includegraphics[width=0.495\textwidth]{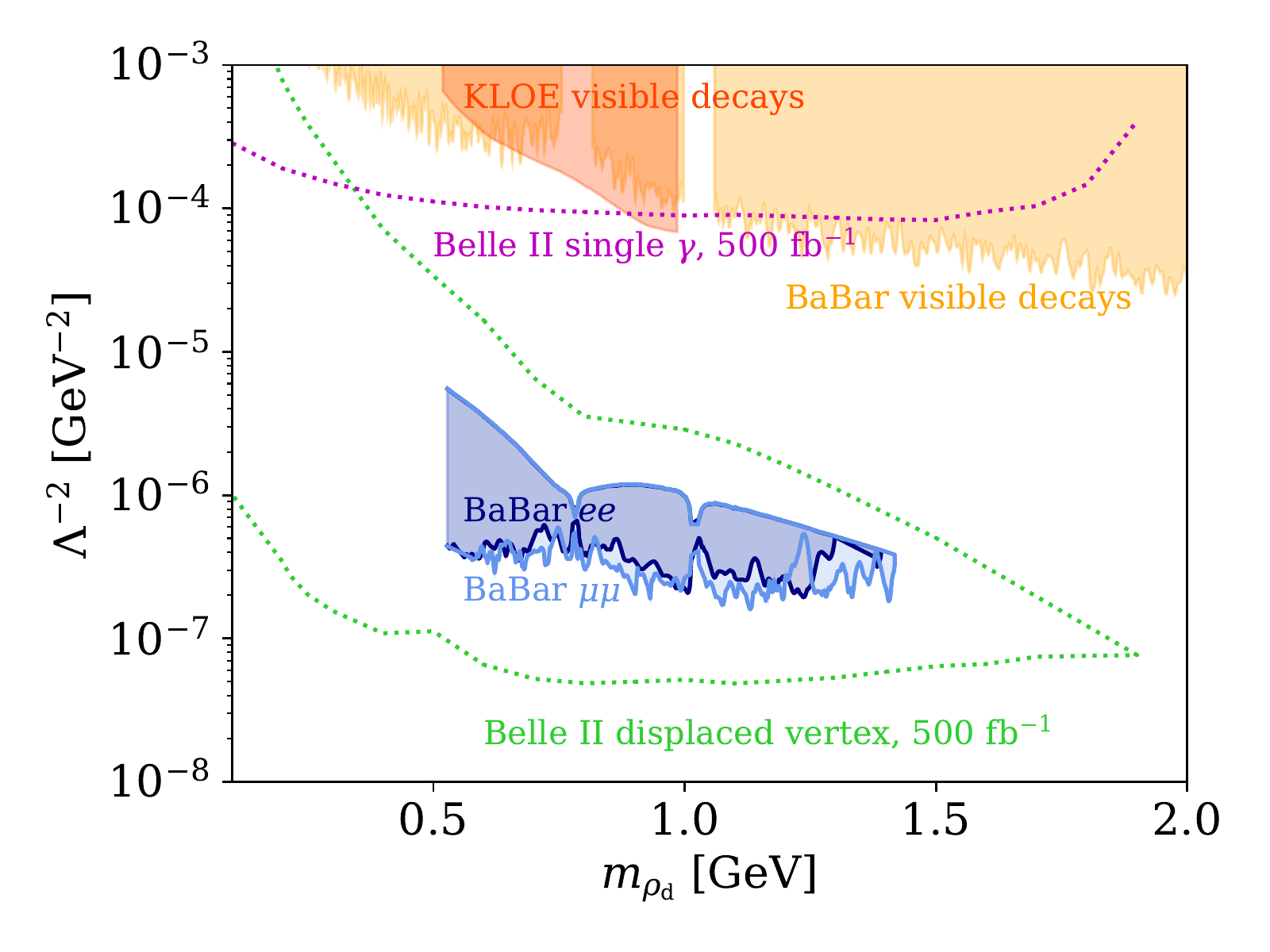}
\includegraphics[width=0.495\textwidth]{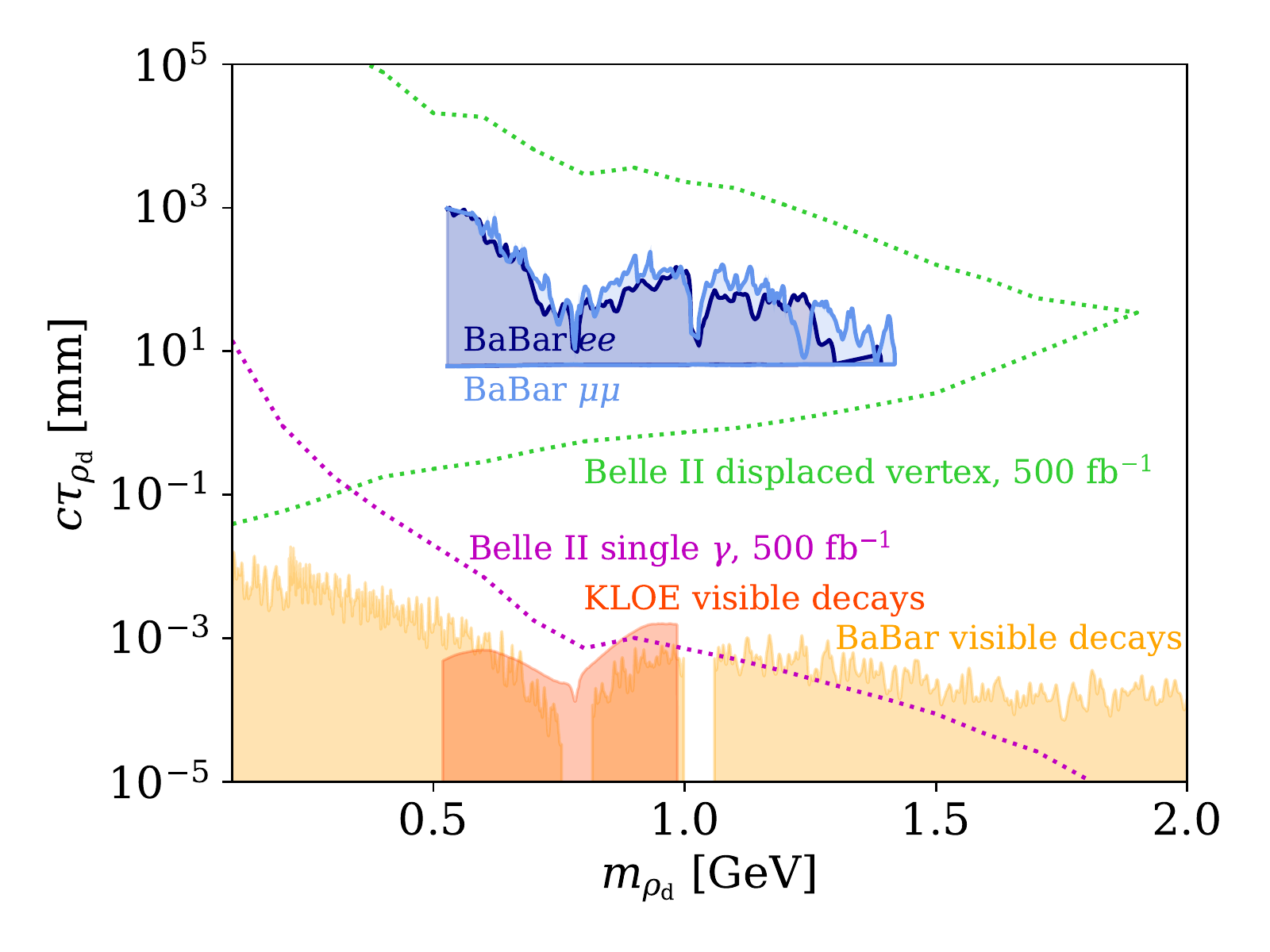}
\caption{Comparison of existing bounds and projected \belletwo sensitivities on the effective production of dark showers with prompt and displaced $\rhodzero$ meson decays as well as for fully invisible dark showers. \label{fig:bfactory_limits}}
\end{center}
\vspace{-5mm}
\end{figure}

A comparison of the different constraints on the low-energy effective theory is shown in figure~\ref{fig:bfactory_limits} in terms of the $\rhodzero$ meson mass and the  effective coupling (left) as well as the $\rhodzero$ meson proper decay length (right). As expected, we find that the proposed single-photon search at \belletwo as well as the searches for visible decays at BaBar and KLOE are sensitive only to relatively large values of $\Lambda^{-2}$, corresponding to very small $\rhodzero$ meson proper decay lengths of the order of \textmu m. 

We find that the sensitivity of the proposed single photon search in terms of $\Lambda$ is approximately constant for $500 \, \mathrm{MeV} \lesssim m_\rhod \lesssim 1.5 \, \mathrm{GeV}$. In this regime the dark shower production cross section is largely independent of $m_\rhod$ (and approximately equal to the perturbative production cross section for a pair of dark quarks of negligible mass) and the multiplicity of $\rhodzero$ mesons is sufficiently small that a large fraction of dark showers remain fully invisible. For comparison, when we simply calculate the perturbative production cross section for a pair of dark quarks following Ref.~\cite{Liang:2021kgw} and assume that the subsequent dark shower always remains invisible, we obtain the almost identical sensitivity estimate $\Lambda^{-2} < 8 \cdot 10^{-5} \, \mathrm{GeV^{-2}}$.\footnote{Note that this bound is significantly weaker than the one in Ref.~\cite{Liang:2021kgw}, because we consider an integrated luminosity of $500\,\mathrm{fb^{-1}}$ rather than $50\,\mathrm{ab^{-1}}$ and do not attempt background subtraction.} For $m_\rhod \lesssim 500\,\mathrm{MeV}$ the sensitivity decreases because there is a greater probability that the dark shower contains visible final states that are vetoed. For $m_\rhod > 1.5 \, \mathrm{GeV}$, on the other hand, the sensitivity is weakened by the requirement $M^2 > 3.5^2 m_\rhod^2$ imposed in our simulation.

The proposed single photon search can improve over existing constraints for $m_\rhod \lesssim 1.1 \, \mathrm{GeV}$. We emphasise that for these $\rhodzero$ meson masses, the \texttt{Pythia} dark shower simulation is expected to be robust, and the requirement $M^2 > 3.5^2 m_\rhod^2$ has no relevant effect on our predictions. Intriguingly, we find that the sensitivities of the proposed single photon search and the proposed displaced vertex search overlap for $m_\rhod \lesssim 330\,\mathrm{MeV}$, such that the entire range of interesting decay lengths from prompt decays to $c\tau \gg 1 \,\mathrm{m}$ can be covered.

\section{Comparison with model-dependent constraints from high-energy accelerators}
\label{sec:highe}

So far we have focused exclusively on experiments that do not have sufficient energy to produce on-shell $Z'$ bosons, such that the resulting constraints depend on the $Z'$ mass and its couplings only through the suppression scale $\Lambda$. In this section we extend our discussion to include also constraints from high-energy accelerators, which break this degeneracy. In particular, while for low-energy experiments the cross section scales proportional to $m_{Z'}^{-4}$, high-energy accelerators achieve a more favourable scaling with increasing $Z'$ mass, such that they possess a clear advantage for the case of TeV-scale $Z'$ bosons~\cite{Bernreuther:2019pfb}. In the following we will instead focus on $Z'$ bosons with a mass below the SM $Z$ boson mass. We will see that in this case constraints from high and low energies are rather complementary, with the details of the comparison depending on the coupling scenario under consideration.

\subsection{Precision observables}

Kinetic mixing between the $Z'$ and the SM hypercharge gauge boson changes the properties of the SM $Z$ boson and thereby affects the Peskin-Takeuchi parameters $S$ and $T$. Following Ref.~\cite{Babu:1997st} we obtain the following expressions:
\begin{align}
    \alpha S & = \frac{4 s_\mathrm{w}^2 \kappa^2}{1-\zeta} \left(1-\frac{s_\mathrm{w}^2}{1-\zeta}\right) \, ,\\
    \alpha T & = \frac{s_\mathrm{w}^2 \kappa^2}{c_\mathrm{w}^2}\frac{1}{1-\zeta} \; ,
\end{align}
where $c_\mathrm{w}$ ($s_\mathrm{w}$) denotes the cosine (sine) of the weak mixing angle, $g$ ($g'$) denotes the weak (hypercharge) gauge coupling and $\zeta = m_{Z'}^2/m_Z^2$.\footnote{Note that Ref.~\cite{Babu:1997st} assumes $\zeta \gg 1$ and therefore approximates $1 - s_\mathrm{w}^2/(1-\zeta) \approx 1$, whereas we are interested in $\zeta < 1$, see also Ref.~\cite{Frandsen:2011cg}.} Using the most recent constraints from Ref.~\cite{ParticleDataGroup:2020ssz}, we obtain the 95\% confidence level upper bound $\kappa < 0.014$ for $m_{Z'} < 60\,\mathrm{GeV}$, slightly stronger than the bounds obtained in Ref.~\cite{Hook:2010tw} using pre-LHC data.\footnote{Note that the confidence level for this bound differs from all the other exclusion limits that we show. The reason is that the SM prediction $S = T = 0$ is in fact excluded by EWPT data at 90\% confidence level, such that we use a more conservative bound instead.}

Another potential constraint arises from the invisible $Z$ boson width, for which any exotic contribution is tightly constrained to be $\Gamma_Z^\text{inv} < 1.5\,\mathrm{MeV}$. Since kinetic mixing induces a small coupling of the SM $Z$ boson to dark quarks, we can translate this constraint into an upper bound on $\kappa e_\mathrm{d}$. Assuming that the dark quarks produced in the $Z$ boson decay remain invisible, one obtains
\begin{equation}
    \Gamma_Z^\text{inv} = N_q \frac{m_Z}{12 \pi} \frac{s_\mathrm{w}^2}{(1-\zeta)^2} \frac{\kappa^2 e_\mathrm{d}^2}{c_\mathrm{w}^2} \; ,
\end{equation}
where $N_q = 6$ denotes the product of dark quark flavours and colours in our set-up. We find that for $e_\mathrm{d} < 1$ and the range of $Z'$ masses that we are interested in, this constraint is weaker than the one from the $S$ and $T$ parameters and we will therefore only consider the latter in the following.

\subsection{Displaced vertex searches at LHCb}

LHCb has shown impressive performance in the search for various types of light LLPs decaying into jets~\cite{LHCb:2014jgs} and leptons~\cite{LHCb:2019vmc}, such as dark photons~\cite{LHCb:2017trq} or axion-like particles~\cite{LHCb:2015nkv,LHCb:2016awg}. While ATLAS and CMS offer the possibility to use the missing transverse momentum of an event to suppress background, LHCb has a clear advantage when it comes to trigger thresholds and the reconstruction of light and relatively soft LLPs, as expected in the case of dark showers with GeV-scale hadrons~\cite{Pierce:2017taw}. Of particular interest in the context of the present work is an inclusive search for di-muon resonances~\cite{LHCb:2020ysn}, for which results have been presented in a model-independent way that allows for a straight-forward reinterpretation.\footnote{In fact, Ref.~\cite{LHCb:2020ysn} also provides an interpretation of these constraints in the context of strongly interacting dark sectors, but the assumed hadron multiplicities do not agree with the predictions of the model that we consider so that a new analysis is needed (see also Ref.~\cite{Cheng:2021kjg}).}

\begin{table}[t!]
    \caption{\label{tab:cuts} Selection cuts of the LHCb search for promptly produced LLPs decaying into a muon pair. Here $\alpha$ denotes the opening angle and $d_T$ denotes the transverse decay length. Note that we assume that the requirement on the number of charged particles from the primary vertex is always satisfied and that the decay topology is always consistent with a promptly produced $\rhodzero$ meson.}\vspace{-5mm}
    \begin{center}
      \begin{tabular}{ccc}
        \toprule
        Cuts on each muon & Cuts on each muon pair & Cuts on the reconstructed LLP \\
        \midrule
        $p_T(\mu) > 0.5\,\mathrm{GeV}$ & $\sqrt{p_T(\mu^+) p_T(\mu^-)} > 1\,\mathrm{GeV}$ & $2\,\mathrm{GeV} < p_T(\rho_\mathrm{d}) < 10\,\mathrm{GeV}$ \\ 
        $2 < \eta(\mu) < 4.5$ & $\alpha(\mu^+\mu^-) >3$\,mrad & $2 < \eta(\rho_\mathrm{d}) < 4.5$ \\
        $10 \,\mathrm{GeV} < p(\mu) < 1000 \,\mathrm{GeV}$ & &  $12\,\mathrm{mm} < d_T(\rho_\mathrm{d}) < 30\,\mathrm{mm}$ \\
        \bottomrule
      \end{tabular}
  \end{center}
\end{table}

Based on an integrated luminosity of $5.1\,\mathrm{fb^{-1}}$, the LHCb analysis considers muon pairs from a displaced vertex with an invariant mass in the range from 0.21 to 3 GeV. The level of background depends decisively on whether or not the inferred momentum of the LLP is required to point back to the interaction point, i.e.\ whether or not the LLP is promptly produced. Since the $\rhodzero$ meson is the only meta-stable particle in our setup, the requirement of prompt production will always be satisfied and we can apply the more restrictive selection requirements. 
Since no significant excess is found in the data, the LHCb analysis places an upper bound on the $\rhodzero$ meson production cross section as a function of its mass for three different $p_T$ bins in the range $2 \, \mathrm{GeV} \leq p_T(\rho_\mathrm{d}) \leq 10 \, \mathrm{GeV}$.

\begin{figure}[t]
\begin{center}
\includegraphics[width=0.495\textwidth]{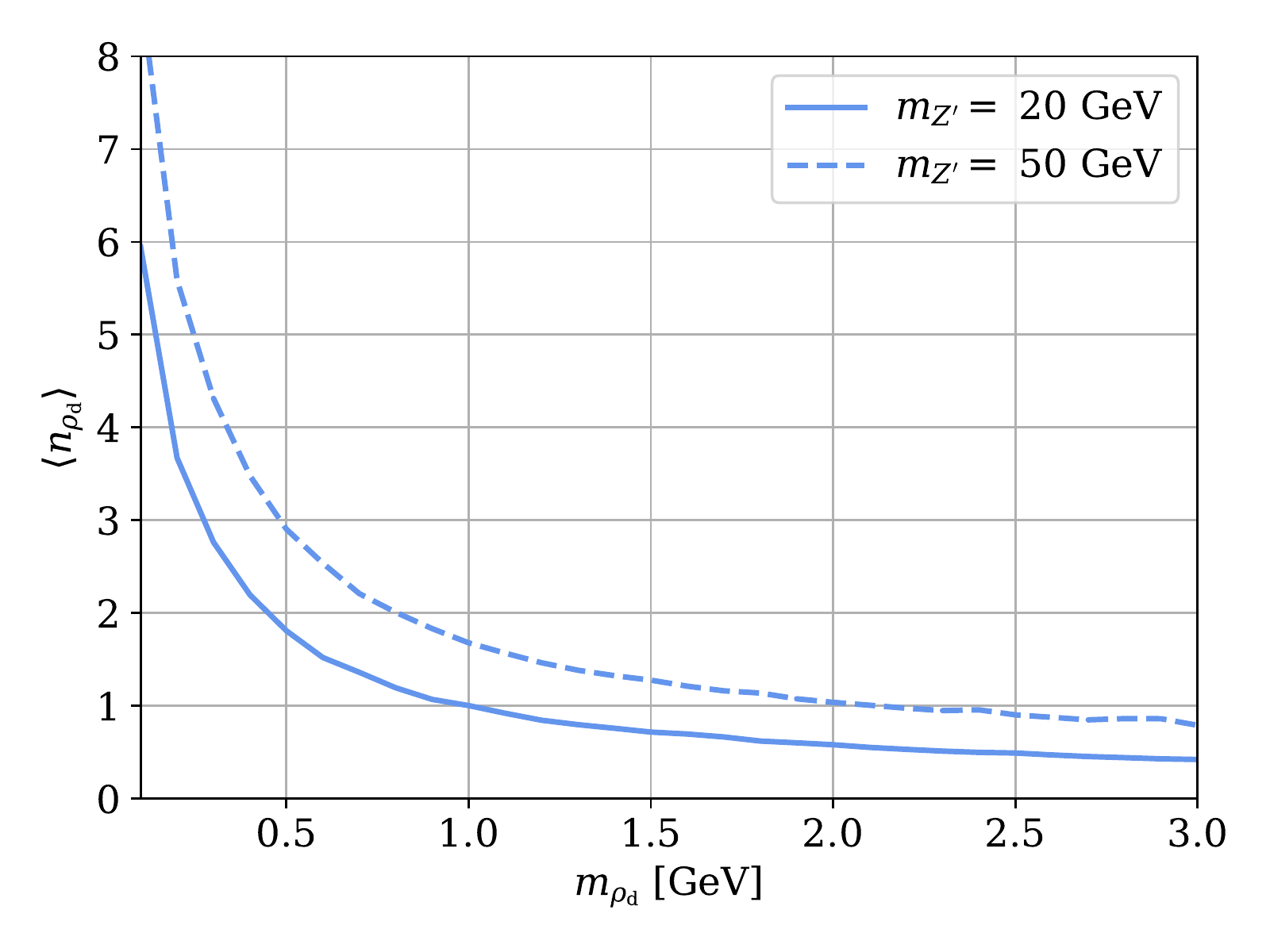}
\includegraphics[width=0.495\textwidth]{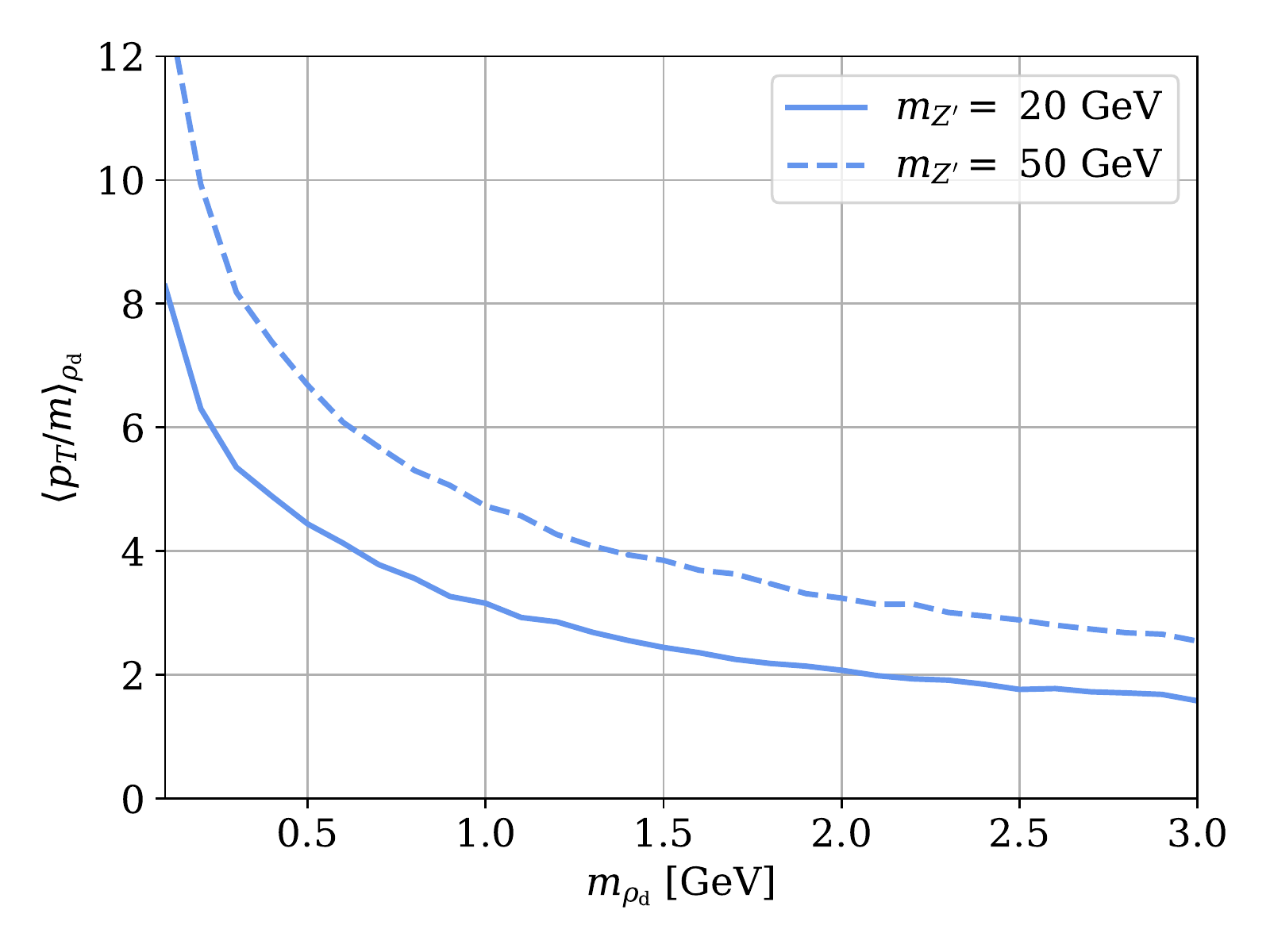}
\caption{\label{fig:lhcb_llp_mult_boost} Average multiplicity (left) and average transverse boost $p_T / m_{\rho_\mathrm{d}}$ (right) of $\rhodzero$ mesons in dark showers produced at the LHC as a function of $m_{\rho_\mathrm{d}}$ for different $Z'$ masses. }
\end{center}
\vspace{-5mm}
\end{figure}

To generate events we simulate the hard process $pp\to q_\mathrm{d} \bar{q}_\mathrm{d}$ using \texttt{MadGraph5} with \texttt{LHAPDF6} parton distribution functions~\cite{Buckley:2014ana}. We allow for an additional hard jet using MLM matching to avoid double counting. The simulation of the dark shower is done with \texttt{Pythia} as described above. After passing the hadronised events to \texttt{DELPHES 3} for detector simulation~\cite{deFavereau:2013fsa}, we apply the selection cuts listed in table~\ref{tab:cuts}. We point out that it is numerically inefficient to directly apply the cut on the transverse decay length $d_T$. Instead, it is more convenient to weight each event with the probability that the decay happens in the sensitive region:
\begin{equation}
    p_\text{decay} = \exp\left(\frac{-d_{T,1}}{c\tau p_T/m_{\rho_\mathrm{d}}}\right) - \exp\left(\frac{-d_{T,2}}{c\tau p_T/m_{\rho_\mathrm{d}}}\right) \; ,
    \label{eq:weights}
\end{equation}
where $d_{T,1} = 12\,\mathrm{mm}$ and $d_{T,2} = 30\,\mathrm{mm}$.

While the masses of the $Z'$ and of the $\rhodzero$ meson affect the simulated distributions in a non-trivial way, a change of couplings only changes the overall cross section and (through the lifetime $\tau$) the event weights. This makes it possible to reuse events generated for a given coupling combination through an appropriate rescaling. Specifically, we use the narrow-width approximation
\begin{equation}
    \sigma(pp\to q_\mathrm{d}\bar{q}_\mathrm{d}) \propto \kappa^2 \text{BR}(Z' \to q_\mathrm{d}\bar{q}_\mathrm{d}) \; ,
    \label{eq:NWA}
\end{equation}
which we have confirmed to be valid for $\Gamma_{Z'} \lesssim m_{Z'}/3$, and recalculate the event weights according to eq.~(\ref{eq:weights}).

For a given value of $m_{Z'}$ we simulate 35 values of $m_{\rho_\mathrm{d}}$ in the range from $0.2\,\mathrm{GeV}$ to $3 \, \mathrm{GeV}$. The average multiplicity and transverse boost of long-lived $\rhodzero$ mesons are shown as functions of their mass $\mrhod$ in figure~\ref{fig:lhcb_llp_mult_boost}. As before, we find that the multiplicity of $\rhodzero$ mesons decreases with increasing mass, which implies an increase of the typical transverse momentum. Conversely, increasing the $Z'$ mass from 20 GeV to 50 GeV leads to an increase in the $\rhodzero$ meson multiplicity, such that the typical transverse momentum increases only slightly and thus remains
within the range covered by the LHCb analysis. We find that the greatest sensitivity is achieved for the highest $p_T$ bin ($5 \, \mathrm{GeV} < p_T(\rho_\mathrm{d}) < 10 \, \mathrm{GeV}$), for which the signal is largest and the background smallest.

\subsection{Results}

\begin{figure}[t]
\begin{center}
\includegraphics[width=0.495\textwidth]{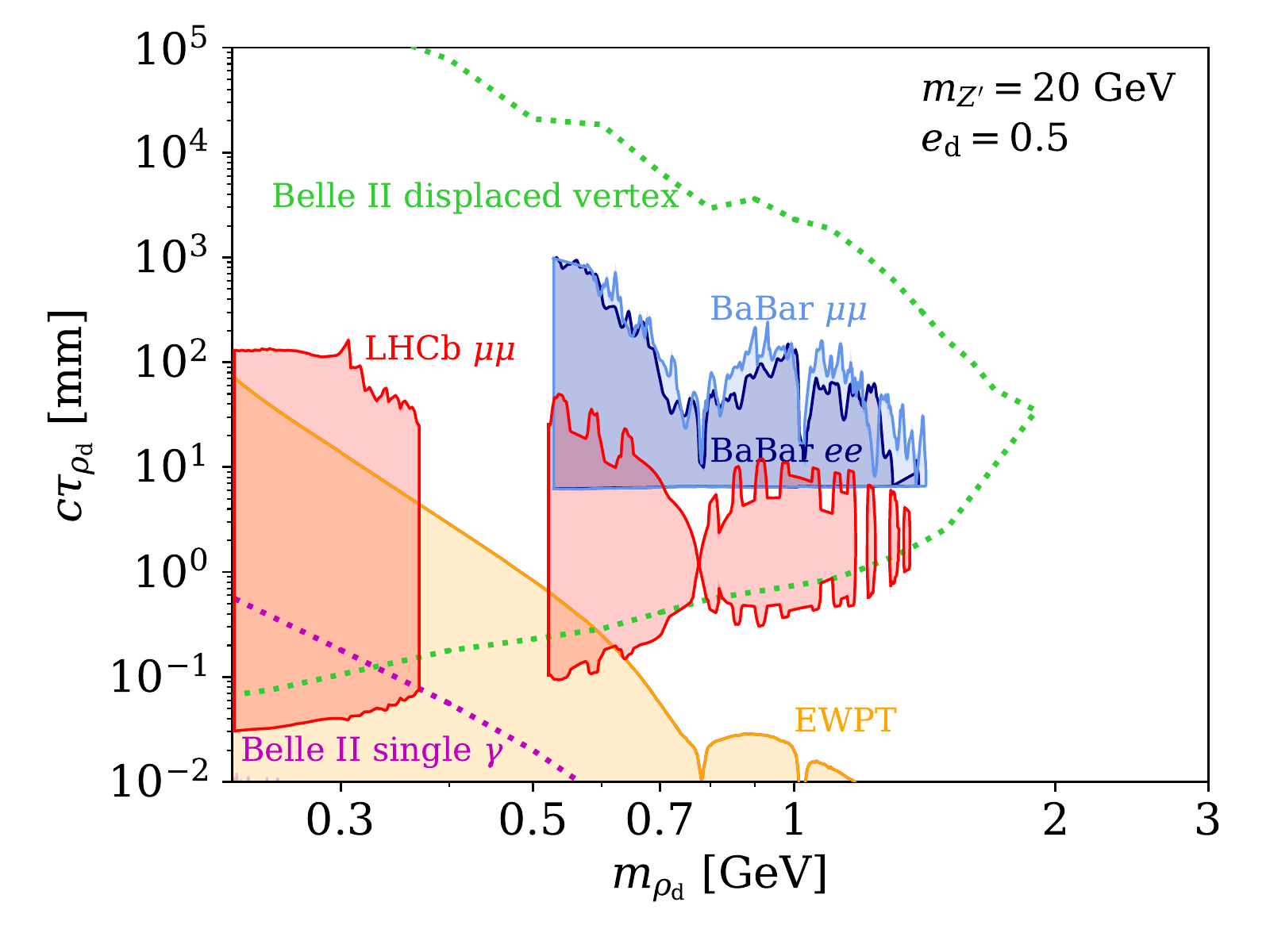}
\includegraphics[width=0.495\textwidth]{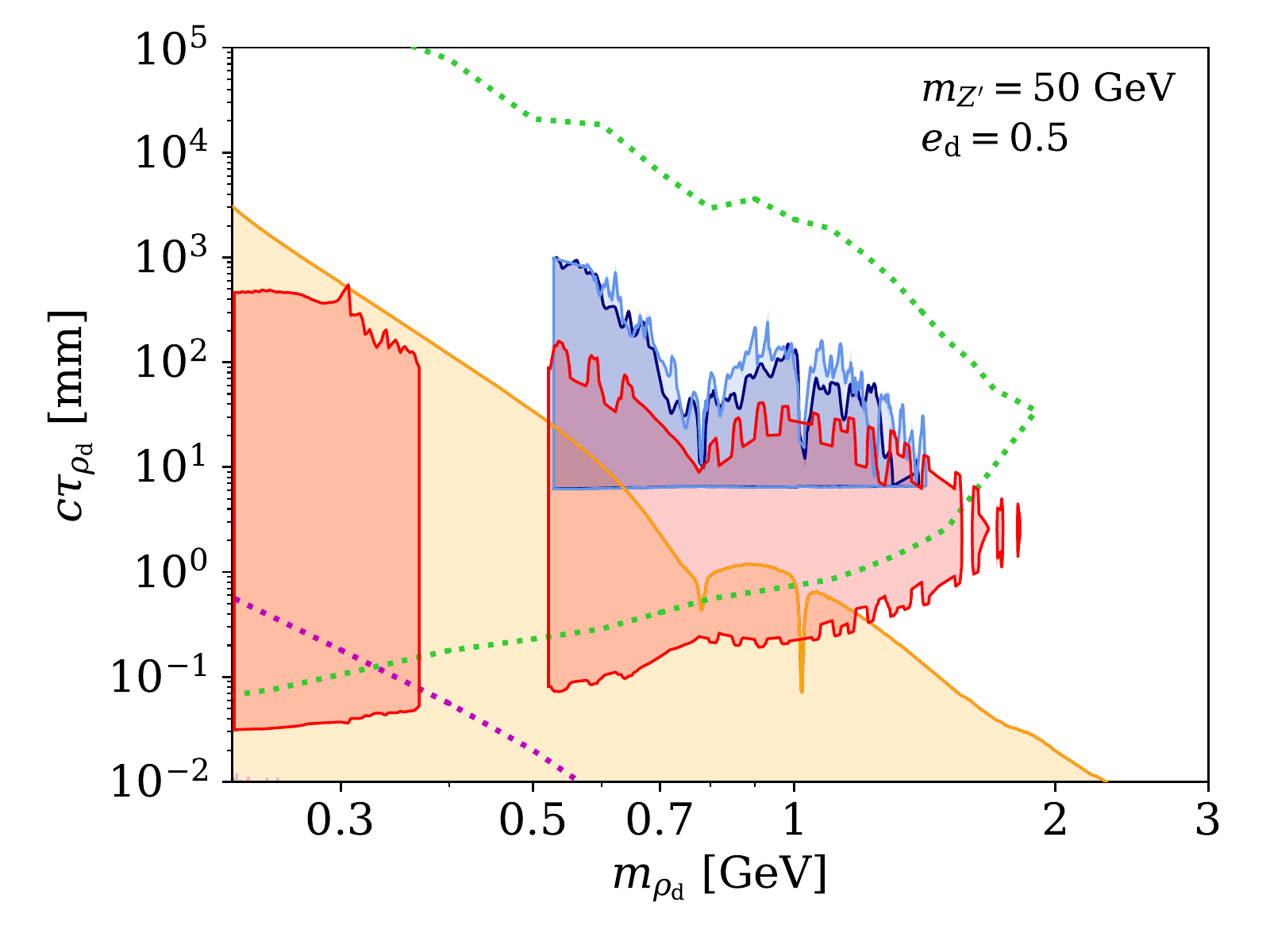}
\includegraphics[width=0.495\textwidth]{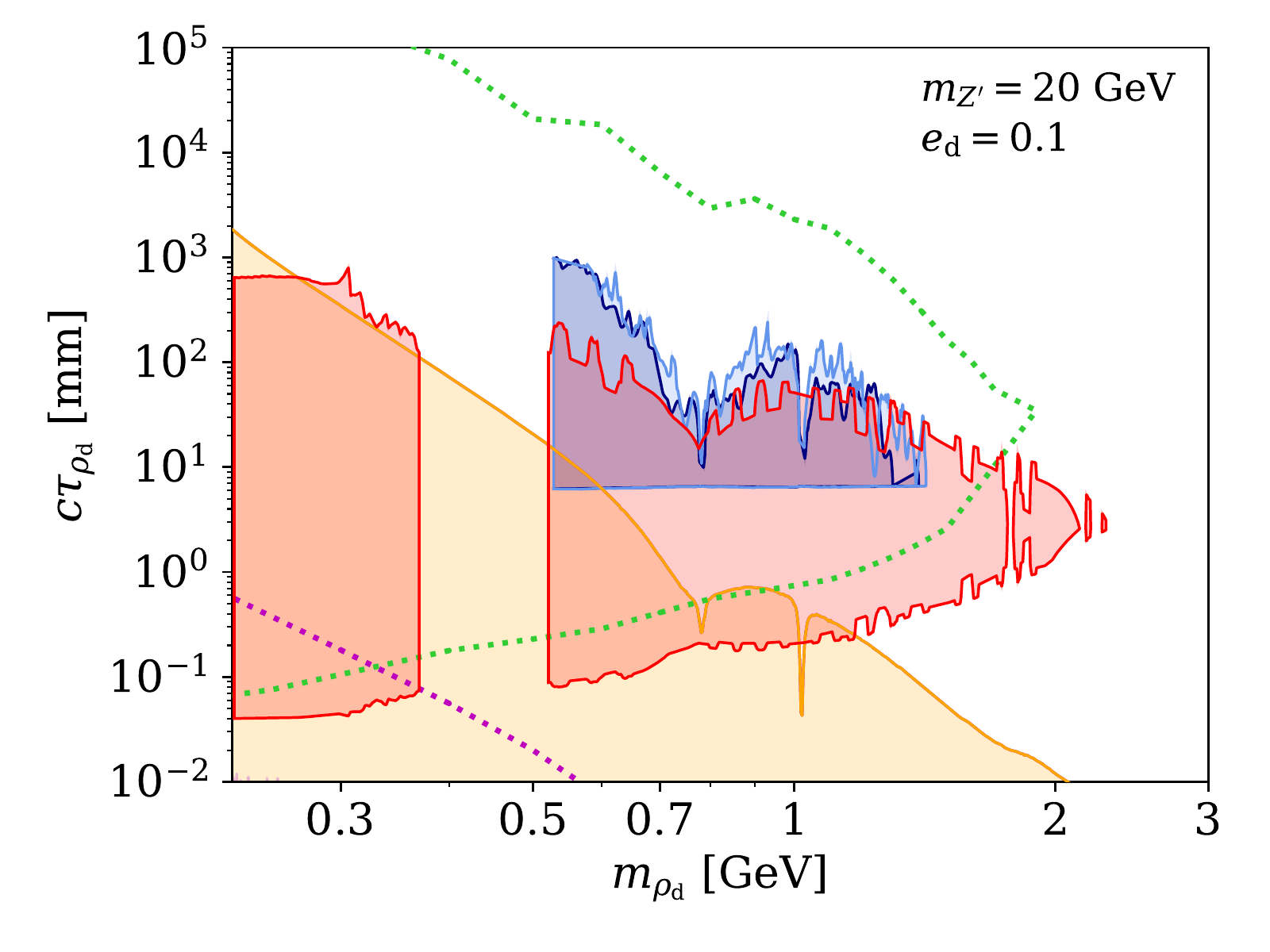}
\includegraphics[width=0.495\textwidth]{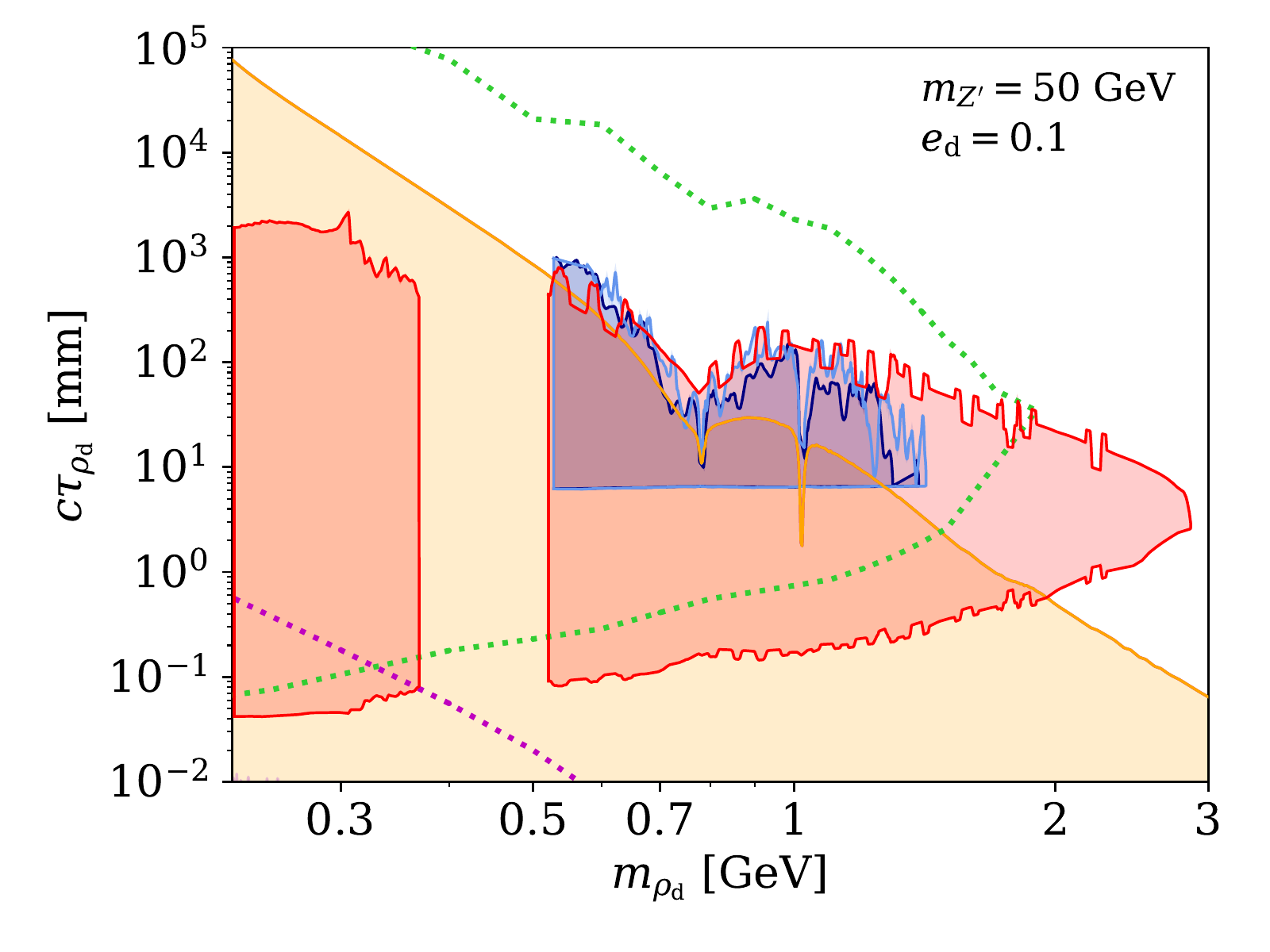}
\includegraphics[width=0.495\textwidth]{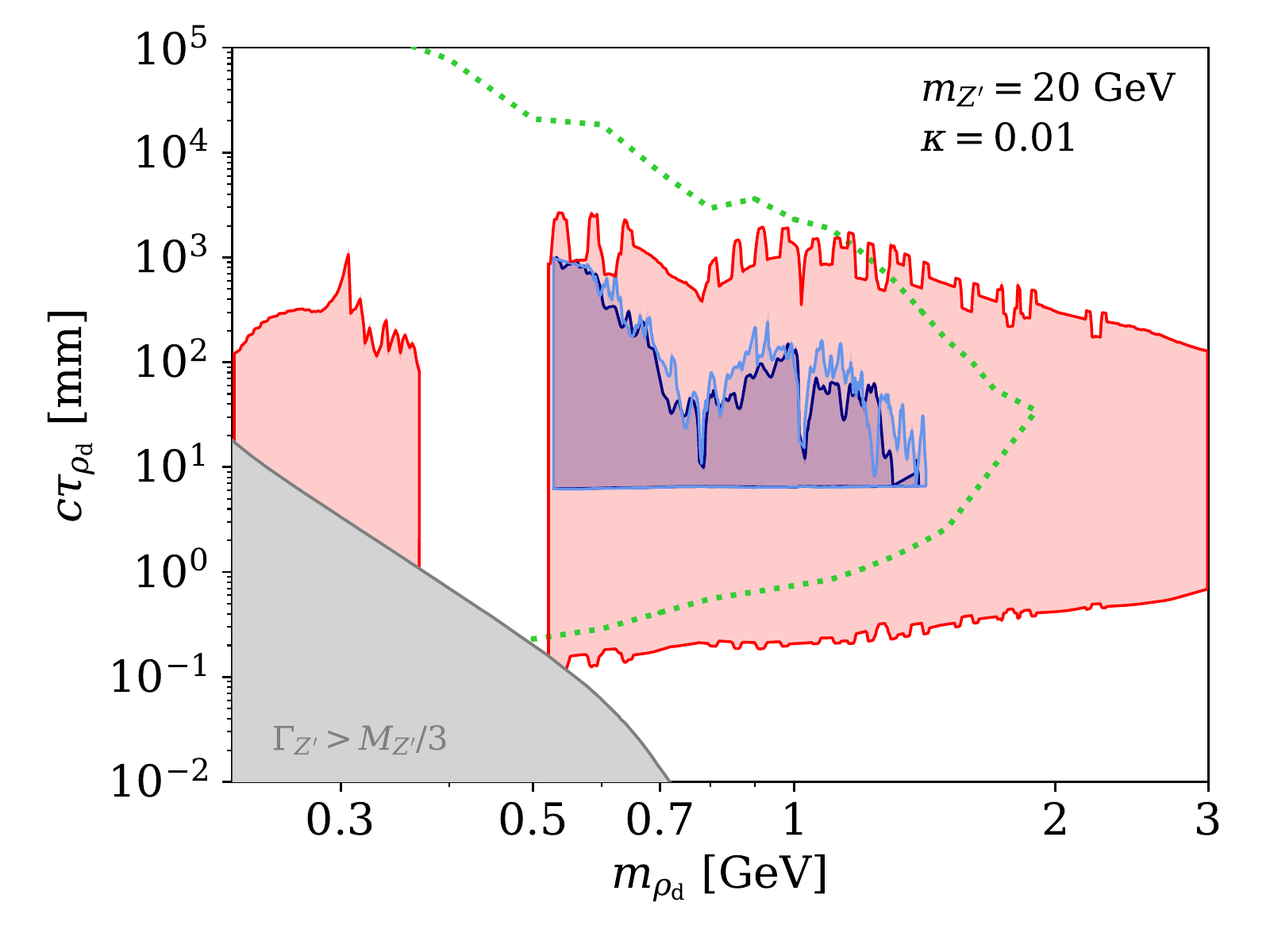}
\includegraphics[width=0.495\textwidth]{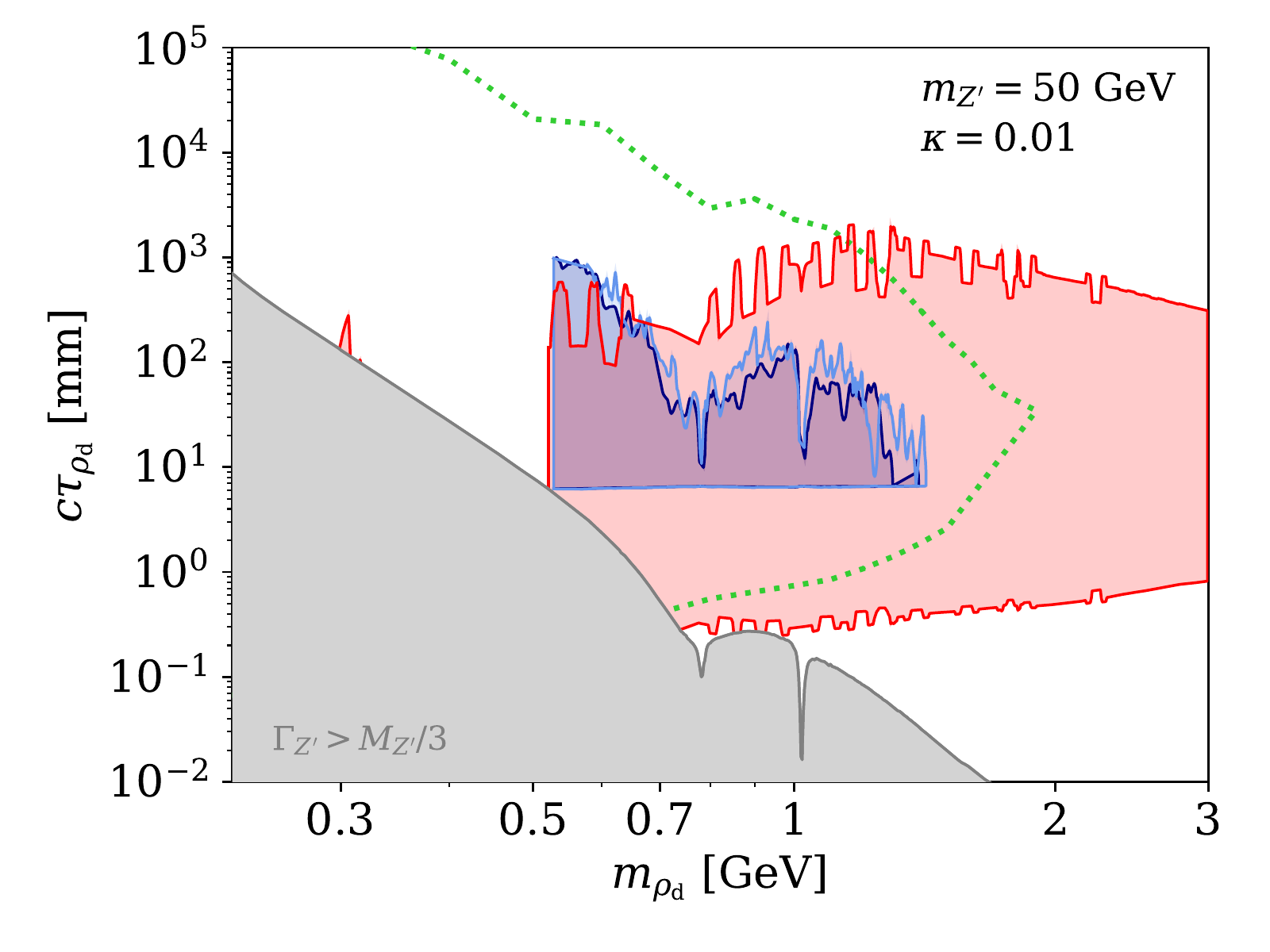}
\caption{\label{fig:results} Constraints in the $m_{\rho_\mathrm{d}}$--$c\tau_{\rho_\mathrm{d}}$ parameter plane for $m_{Z'} = 20 \, \mathrm{GeV}$ (left) and $m_{Z'} = 50 \, \mathrm{GeV}$ (right) and three different coupling scenarios: fixed $e_\mathrm{d} = 0.5$ and variable $\kappa$ (top row), fixed $e_\mathrm{d} = 0.1$ and variable $\kappa$ (middle row), and fixed $\kappa = 0.01$ and variable $e_\mathrm{d}$ (bottom row).}
\end{center}
\vspace{-5mm}
\end{figure}

To highlight the different scaling of constraints from high and low energies, we consider two different values of the $Z'$ mass ($m_{Z'} = 20\,\mathrm{GeV}$ and $m_{Z'} = 50\,\mathrm{GeV}$) and three different coupling scenarios:
\begin{itemize}
    \item Scenario 1: fixed $e_\mathrm{d} = 0.5$, variable $\kappa$;
    \item Scenario 2: fixed $e_\mathrm{d} = 0.1$, variable $\kappa$;
    \item Scenario 3: Fixed $\kappa = 0.01$, variable $e_\mathrm{d}$.
\end{itemize}
In the first two scenarios, there will be non-trivial constraints from EWPT, while the condition $\Gamma_{Z'} < m_{Z'} / 3$ is always satisfied, whereas in the third scenario the EWPT constraints are always satisfied and the requirement on the width needs to be checked explicitly.

Our results are summarised in figure~\ref{fig:results} in the $m_{\rho_\mathrm{d}}$--$c \tau_{\rho_\mathrm{d}}$ parameter plane. We remind the reader that in this parameter plane the constraints from low-energy experiments are independent of the value of the $Z'$ mass or the specific coupling scenario under consideration and simply correspond to the ones shown in figure~\ref{fig:bfactory_limits}. The most striking observation is that there is a clear complementarity between the different constraints: While BaBar and \belletwo are most sensitive to $\rhodzero$ mesons with a proper decay length of 10--100 mm, the sensitivity of LHCb peaks at 1-10 mm. Even smaller decay lengths can be constrained by EWPT and the search strategies discussed in section~\ref{sec:lowe}.

As expected, constraints from high-energy experiments get stronger with increasing $Z'$ mass, even though the effect is more pronounced for the precision observables than for the constraints from LHCb. To understand the behaviour of the LHCb constraint for the different coupling scenarios, we can once again refer to the narrow-width approximation in eq.~(\ref{eq:NWA}). For $e_\mathrm{d} \gg e\kappa$ one finds $\text{BR}(Z' \to q_\mathrm{d}\bar{q}_\mathrm{d}) \approx 1$, such that the production cross section is simply proportional to $\kappa^2$. For given values of $c \tau_{\rho_\mathrm{d}}$ and $m_{\rho_\mathrm{d}}$ the case $e_\mathrm{d} = 0.1$ hence corresponds to larger values of $\kappa$ (and hence larger production cross sections) than for $e_\mathrm{d} = 0.5$, leading to correspondingly stronger constraints from LHCb. The case where we fix $\kappa = 0.01$ corresponds to even smaller values of $e_\mathrm{d}$ and hence even stronger bounds.

Most importantly, we find that for all scenarios that we consider there is a strong case for \belletwo to search for displaced vertices from strongly interacting dark sectors. The unconstrained parameter regions that can be probed with such a search are largest in the case of small $m_{Z'}$ and large $e_\mathrm{d}$, but even in less ideal scenarios the reach of LHCb does not extend beyond decay lengths of about $10^3$ mm, well below what may be achieved with \belletwo.

\section{Conclusions}

Strongly interacting dark sectors in general and dark showers at accelerators in particular have received rapidly growing interest in recent years. In the present work we have considered an effective theory of dark shower production at low-energy experiments such as $B$ factories. This description is valid if the centre-of-mass energy of the experiment is small compared to the mass of the mediator that couples the dark and the visible sector, but large compared to the confinement scale of the dark sector. Under these conditions, we find that both the dark shower production cross section and the lifetime of the long-lived $\rhodzero$ mesons can be expressed as a function of the $\rhodzero$ meson mass $m_\rhod$ and the suppression scale $\Lambda$ only. This makes it possible to directly compare a range of low-energy experiments in terms of these two parameters.

Specifically, we have compared the exclusion limits from an existing DV search at BaBar and from searches for prompt $\rhodzero$ meson decays at BaBar and KLOE with the expected sensitivity of two proposed \belletwo searches: First, a DV search, which uses a range of different triggers to target LLPs within the dark shower decaying into $e^+e^-$, $\mu^+ \mu^-$, $\pi^+ \pi^-$ or $K^+ K^-$.  Second, a single-photon search, which probes the case that the dark shower remains fully invisible. We find that the former promises substantial improvements over existing exclusion limits, in particular for long lifetimes, already with the currently collected amount of data. The latter search, on the other hand, suffers from the complication that the invariant mass of the dark shower is not fixed, and hence the photon from initial state radiation is not mono-energetic. This makes it much more difficult to distinguish signal and background and limits the sensitivity of the search unless background subtraction becomes possible. Nevertheless, even for our conservative estimates we expect some improvement over existing bounds from such an analysis.

Whenever considering an effective theory, it is essential to explore possible high-energy completions and understand whether they are phenomenologically viable. For this purpose we have considered the case that the effective interaction arises from a $Z'$ mediator with a mass below the $Z$ boson mass and kinetic mixing with the SM hypercharge field. The couplings of such a $Z'$ mediator are tightly constrained by EWPT, and we update the corresponding bounds. Furthermore, we consider dark shower production at the LHC in order to obtain exclusion limits from a model-independent search for LLPs at LHCb. We find that this search is highly sensitive to our model, but targets somewhat smaller $\rhodzero$ meson lifetimes than what can be probed by \belletwo. Moreover, the strength of the LHCb constraints depends in a non-trivial way on the assumed coupling structure and the $Z'$ mass. This model dependence highlights the complementarity of experiments operating at different energies.

The set-up that we have studied can potentially be applied to a much wider range of experiments. Additional constraints on the low-energy effective theory may be obtained from fixed-target experiments, in particular those with large angular coverage like SHiP~\cite{SHiP:2015vad} and/or off-axis detectors like SHADOWS~\cite{Baldini:2021hfw}. In the context of the LHC, it will be interesting to study the sensitivity of proposed experiments dedicated to the search for LLPs, such as CODEX-b~\cite{Aielli:2019ivi}, ANUBIS~\cite{Bauer:2019vqk} or MATHUSLA~\cite{MATHUSLA:2018bqv,MATHUSLA:2020uve} as well as to the search for forward physics such as FASER~\cite{FASER:2019aik} and other experiments at a Forward Physics Facility~\cite{Anchordoqui:2021ghd} (see the recent study in Ref.~\cite{Archer-Smith:2021ntx}). Nevertheless, we emphasise that dark shower production is not strongly peaked in the forward direction, making \belletwo particularly well-suited to probe this model.

Furthermore, if the dark pions are completely stable and constitute a significant fraction of the DM abundance of the universe, additional constraints are expected from direct detection experiments, which can search for the scattering of dark pions off nuclei or electrons via the effective interaction given in eq.~(\ref{eq:smfermions_darkquarks_effective}). For dark pions at the GeV scale, these constraints are however fairly weak and only probe the parameter space corresponding to prompt decays of the $\rhodzero$ mesons.  Nevertheless, substantial sensitivity improvements are expected in coming years, offering the opportunity to observe signals from strongly interacting dark sectors across several experiments.

To conclude, let us emphasise that the various analyses considered in the present work do not make use of the full complexity of dark showers in the sense that we assume that most of the dark shower remains invisible and only a single DV is detected in a given event. While this approach is appropriate if most of the light dark mesons are stable, it is equally conceivable that most (or even all) of the light dark mesons can decay into SM final states, leading to a multitude of DVs. While this should make it easier in principle to trigger on such events and reject backgrounds, dedicated analysis strategies will be needed to make the most of the enormous potential of ongoing and near-future accelerators to probe strongly interacting dark sectors.

\acknowledgments

We thank Michael Kr{\"a}mer, Brian Shuve and Susanne Westhoff for discussions. This work is funded by the Deutsche Forschungsgemeinschaft (DFG) through the Collaborative Research Center TRR 257 ``Particle Physics Phenomenology after the Higgs Discovery'' under Grant 396021762 – TRR 257, the Emmy Noether Grant No. KA 4662/1-1 and Germany’s Excellence
Strategy -- EXC 2121 ``Quantum Universe'' -- 390833306, through the Helmholtz (HGF) Young Investigators Grant No.\ VH-NG-1303, and through the Natural Sciences and Engineering Research Council of Canada, Compute Canada and CANARIE. This manuscript has been authored by Fermi Research Alliance, LLC under Contract No. DE-AC02-07CH11359 with the U.S. Department of Energy, Office of Science, Office of High Energy Physics.

\bibliographystyle{JHEP_improved}
\bibliography{biblio}

\end{document}